\documentclass[sigconf, 9pt]{acmart}
\newcommand{\sssec}[1]{\vspace*{0.05in}\noindent\textbf{#1}}

\newcommand{\name}{\emph{ToMoBrush}}
\newcommand{\swarun}[1]{\textcolor{black}{#1}}
\newcommand{\kuang}[1]{\textcolor{black}{#1}}
\newcommand{\rr}[1]{\textcolor{black}{#1}}
\newcommand{\feb}[1]{\textcolor{black}{#1}}
\usepackage{graphicx, subcaption, amsmath,mathrsfs}
\usepackage[capitalize]{cleveref}
\setcopyright{acmcopyright}
\setcopyright{none}
\renewcommand\footnotetextcopyrightpermission[1]{}
\copyrightyear{2018}
\acmYear{2018}
\acmDOI{XXXXXXX.XXXXXXX}
\acmConference[]{}{0}{0} 
\title{ToMoBrush: Exploring Dental Health Sensing \\ using a Sonic Toothbrush}
\author{Kuang Yuan}
\affiliation{Carnegie Mellon University\country{}}
\email{kuangy@cmu.edu}
\author{Mohamed Ibrahim}
\affiliation{Carnegie Mellon University\country{}}
\affiliation{Hewlett Packard Labs\country{}}
\email{ibrahim@hpe.com}
\author{Yiwen Song}
\affiliation{Carnegie Mellon University\country{}}
\email{yiwens2@andrew.cmu.edu}
\author{Guoxiang Deng}
\affiliation{Carnegie Mellon University\country{}}
\email{gdeng@andrew.cmu.edu}
\author{Suvendra Vijayan}
\affiliation{School of Dental Medicine\\ University of Pittsburgh\country{}}
\email{suv16@pitt.edu}
\author{Robert Nerone}
\affiliation{School of Dental Medicine\\ University of Pittsburgh\country{}}
\email{ron24@pitt.edu}
\author{Akshay Gadre}
\affiliation{University of Washington\country{}}
\email{gadre@uw.edu}
\author{Swarun Kumar}
\affiliation{Carnegie Mellon University\country{}}
\email{swarun@cmu.edu}

\begin{document}
\begin{abstract}

Early detection of dental disease is crucial to prevent adverse outcomes. Today, dental X-rays are currently the most accurate gold standard for dental disease detection. Unfortunately, regular X-ray exam is still a privilege for billions of people around the world. In this paper, we ask: "Can we develop a low-cost sensing system that enables dental self-examination in the comfort of one's home?"

This paper presents \name, a dental health sensing system that explores using off-the-shelf sonic toothbrushes for dental condition detection. Our solution leverages the fact that a sonic toothbrush produces rich acoustic signals when in contact with teeth, which contain important information about each tooth's status. \name\ extracts tooth resonance signatures from the acoustic signals to characterize varied dental health conditions of the teeth. We evaluate \name\ on 19 participants and dental-standard models for detecting common dental problems including caries, calculus, and food impaction, achieving a detection ROC-AUC of 0.90, 0.83, and 0.88 respectively. \kuang{Interviews with dental experts validate \name's potential in enhancing at-home dental healthcare.}

\end{abstract}
% \begin{CCSXML}
% <ccs2012>
%    <concept>
%        <concept_id>10003120.10003138</concept_id>
%        <concept_desc>Human-centered computing~Ubiquitous and mobile computing</concept_desc>
%        <concept_significance>500</concept_significance>
%        </concept>
%    <concept>
%        <concept_id>10010405.10010444.10010447</concept_id>
%        <concept_desc>Applied computing~Health care information systems</concept_desc>
%        <concept_significance>300</concept_significance>
%        </concept>
%  </ccs2012>
% \end{CCSXML}

% \ccsdesc[500]{Human-centered computing~Ubiquitous and mobile computing}
% \ccsdesc[300]{Applied computing~Health care information systems}

\maketitle
\pagestyle{plain}

 \section{Introduction}
 
% \begin{center}
%   \vspace*{0.1in}\textit{For there was never yet a philosopher \\ that could endure the toothache patiently. \\ --  William Shakespeare, ‘Much Ado About Nothing’.}
% \end{center}

Dental disease is a major public health challenge, that can cause pain and infections which may lead to problems with eating, speaking, and even social interaction~\cite{cdc}. Early detection of dental disease is crucial, since it enables interventions that can halt disease progression and prevent the onset of adverse outcomes~\cite {deep2000screening}. Due to the lack of awareness and availability of dental health monitoring, over \$45 billion in US productivity is lost each year due to untreated dental disease~\cite{cdc}. Today, dental X-rays in dental clinics are the most accurate gold standard for dental disease detection~\cite{x-ray}.  Unfortunately, access to dental X-ray diagnostics \rr{at sufficient regularity} is still a privilege for billions of people around the world, due to limited availability, high cost, \rr{and lack of awareness}. In this paper, we explore a low-cost solution for dental health monitoring that users can use regularly in the comfort of their homes. 
% We believe such a solution can prove invaluable to identify potential dental problems early, even for those with access to care.
\rr{We believe such a system can supplement the dental healthcare system even for those with access to professional dental care, by providing early warnings of potential issues in between the dental visits. }
% More that XXX million people around the world visit the dentist every year for problems related to tooth decay and breakages. Today, a dentist in a well-equipped dental care center, typically performs an X-ray exam of the patients teeth for identifying these problems and remedying them. Unfortunately, access to a dentist with an X-ray machine for performing such an exam is a privilege that over XXX billion people around the world cannot afford due to availability and cost. Dentists without access to the above equipment typically rely on visual aids provided by simple tools to detect problems on the surface of teeth. However, studies have shown that this approach of visual diagnosis typically fails to detect YYY \% of decay and breakages hidden inside the tooth leading to further complications. Thus, it is critical to develop a low-cost alternative for detect these hidden breakages and decay of teeth that can overcome the limitations of such a visual approach.

While there is rich prior work on dental health sensing using X-rays, advanced cameras or other infrastructure~\cite{sharma2022ph,hibst2001detection}, these are designed for clinical settings and are expensive. The few systems designed for at-home dental sensing largely focus on toothbrush localization to study brushing habits~\cite{MET,mTeeth,mOral,ToothbrushSound,luo2018brush, huang2016}. 
Perhaps the closest work targeting low-cost dental health sensing is one prior system that uses smartphone cameras for dental self-exams~\cite{OralCam}.
% Perhaps closest to our work is one prior system that uses smartphone cameras for dental self-exams~\cite{OralCam}.
However, this solution can only detect visually diagnosable diseases such as surface cavities and struggles with low-light conditions inside the mouth.

\begin{figure}
    \centering
    % \vspace{0.1in}
    \includegraphics[width=\linewidth]{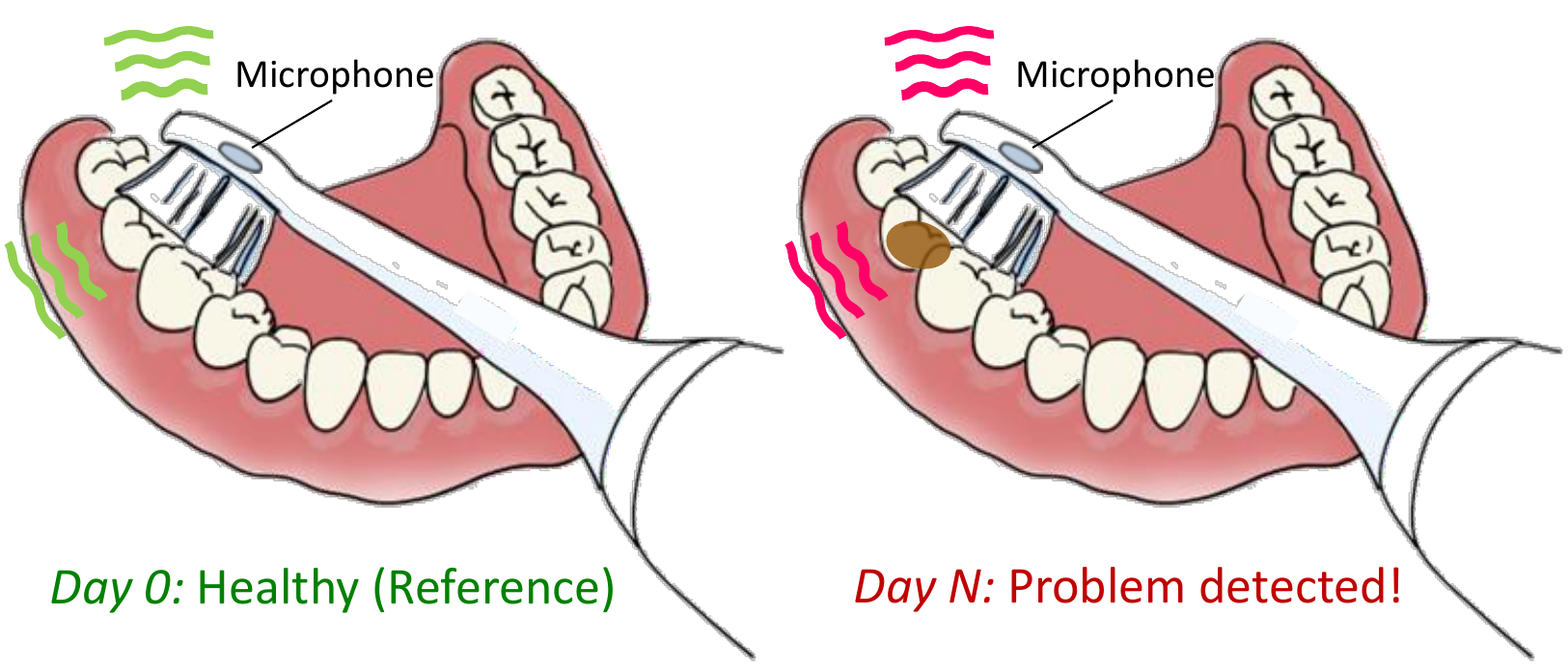}
    \vspace{-0.25in}
    \caption{\name\ is a dental health sensing tool composed of a commodity sonic toothbrush with a microphone integrated close to the brush head. It detects the changes in the resonances of teeth to enable dental health sensing.}
    \vspace{-0.1in}
    \label{fig:device}
\end{figure}

This paper presents \name\footnote{Tooth Monitoring Brush / Tomorrow's Toothbrush}, which explores the potential of using a commodity electric sonic toothbrush \rr{with a minimum hardware modification} for dental health sensing to enable regular, at-home dental \sloppy self-examination. Rather than viewing a toothbrush purely as a cleaning instrument, \name\ leverages the fact that an electric toothbrush is also a rich source of acoustic signals, that are generated by rapid automatic bristle vibrations. When the brush is in contact with a tooth, the tooth also vibrates along with the toothbrush and produces distinct acoustic signals depending on the tooth's state.

% Fig.~\ref{fig:device} visualizes \name's overall setup and architecture. The \name\ platform is composed of a microphone bound to the back of a sonic toothbrush that records the audio when the toothbrush is on. Our system attempts to extract an acoustic signature of every tooth inside the user's mouth. We then show how we can identify and extract teeth deviation from these signatures recorded by the user daily, and alert the toothbrush-user about the potential problematic teeth locations, which necessitate potential follow-up with a dentist. Our benchmark evaluation on dental-standard teeth models achieves a XXX\% accuracy in isolating the signature of individual teeth as well as a XX average classification sensitivity in detecting dental caries, calculus, and stuck food, which are three of the most common dental conditions in the population. Further, we perform an IRB-approved user study in a dental clinical center on XXX patients and demonstrate the efficacy of detecting XX dental caries and XX dental calculus
\kuang{Fig.~\ref{fig:device} visualizes the idea of \name. \rr{An external} microphone is integrated close to the brush head of an off-the-shelf sonic toothbrush, which captures the acoustic signals when the toothbrush is on. Whenever the user wants to perform an at-home dental self-exam, \name\ instructs the user to brush their teeth one at a time along a pre-specified video-guided pattern. } \swarun{We note that this video-guided brushing procedure is performed in addition to regular oral hygiene and does not involve toothpaste. } \kuang{During the brushing procedure, \rr{the microphone on} \name\ captures the acoustic signals from each tooth of the user to detect changes in the tooth's status. }\swarun{These signals are compared with reference signals collected from the patient's healthy teeth (post any treatment) at or shortly after the patient's most recent dental visit. } \kuang{Once a significant deviation is detected compared with the reference signals, \name\ alerts the user about the potential disease with the tooth location, which enables an early follow-up with a dentist. 
We collaborated with dental experts to identify three common dental conditions that \name\ can potentially detect at home and provides support to prevent more severe diseases, including dental caries (cavities from tooth decay), calculus (hardened plaque), and interdental food impaction (food between teeth). 
We perform an IRB-approved user study in a dental clinical center and in a home setting on 19 participants. \name\ achieves a 0.87 average area under receiver operating characteristic curve (ROC-AUC) value for detecting these three dental conditions. We note that for caries and calculus detection, \name's accuracy is comparable to state-of-the-art camera-based systems~\cite{OralCam}.  
}

\name's approach relies on a key observation: Though every sonic toothbrush is designed to vibrate at some specific frequency (usually in 200-400~Hz) for the purpose of teeth cleaning, it also generates harmonics in higher frequencies up to around 20~kHz, due to the imperfections in the vibration motor. As a result, when the toothbrush is in contact with the teeth, it essentially serves as a wideband excitation signal source. The vibration from the toothbrush head traverses and reverberates across the teeth and bones, resulting in a frequency-dependent resonant response, which is unique to the shape, topology, as well as material of the teeth. Thus, when teeth experience health concerns, such as caries, calculus, or even food impaction, this acoustic resonance behavior changes accordingly. \name\ seeks to analyze the difference in the resonance behavior to enable dental health sensing. To achieve this, \name\ addresses two major challenges:

% \name's approach relies on the key observation that most electric tooth-brushes generate sound when turned on and these signals also travel through the teeth when brushing the teeth. While in isolation, these signals are perfect harmonic tones (easier to characterize), upon impact of the teeth with the bristles of the tooth brush, these same signals spread across a wide-bandwidth across the acoustic spectrum (better resolution for sensing). \name\ attempts to extract the health behavior of the teeth that the toothbrush is touching by analyzing this wideband response of the signal from the mouth. The rest of this paper deals with the two major challenges in building such a system: (1) Building an acoustic signature from toothbrush response that captures the behavior of teeth, and (2) Making the system robust to human factors (such as orientation and speed of brushing).

% The resonance behavior ($H(f)$) of the teeth which we desired can be modeled as a linear system: Received signal $Y(f)=H(f)S(f)$, where $S(f)$ is the excitation signal from the toothbrush. Most previous sensing systems for resonance behaviors of object~\cite{milton} have the control to the input excitation signal, so that $S(f)$ can be divided out from the received signal to get clean resonance features.  However, we don't have the access to control the

\sssec{Extracting and Analyzing a Tooth Resonance Signature:} \name\ seeks to model the resonance behavior of the teeth when in contact with the excitation signal from a toothbrush. Indeed, in classical vibrometry, where one models the resonant behavior of objects (e.g. rail lines~\cite{clark2004rail}, rotten fruit~\cite{lee2013evaluation,schotte1999acoustic}, cracked objects~\cite{industryobjectcrackanalysis}) by exciting them with acoustic signals, one has a firm knowledge of the precise acoustic signal applied to the object. However, this is not the case for our system -- we have no ability to control the vibration of an off-the-shelf electric toothbrush. 
Even worse, the toothbrush's vibration behavior is not stable during the brushing procedure. In contrast to static vibrations, brushing the teeth can also change the vibration pattern of the toothbrush. Furthermore, different strength of brushing can also induces different changes. This rules out a one-time calibration where one records the acoustic signal from the toothbrush before any brushing occurs.
% Even worse, the brush's vibration behavior changes every time it is restarted, and it also becomes different when attached to the teeth compared with vibrating alone. This rules out a one-time calibration step where one records the audio signal from the toothbrush before any brushing occurs.  

To tackle this challenge, we carefully model the vibration system including toothbrush, tooth resonance, as well as brushing strength and movement. We propose an algorithm to separate these different factors and extract clean tooth resonance signatures based on a key observation: Though these factors share the same frequency band, their rates of change across the frequencies are different. In Sec.~\ref{sec:signature}, we discuss how we adapt a technique that is widely used in speech processing to separate the glottal excitation and vocal tract resonances. Specifically, we convert the signal into the cepstrum domain where these distinct behaviors are easily separable (detailed in Sec.~\ref{sec:signature}). \kuang{After obtaining the tooth resonance signature, we further develop a feature selection algorithm to select specific regions of signature that are specialized for detecting three different dental conditions (Sec.~\ref{sec:feature}), and perform health detection by comparing the signatures with prior \swarun{healthy reference measurements }(Sec.~\ref{sec:detection}).}
% To tackle this challenge, we leverage a key observation from our experiments: the excitation signal from the brush are noisy harmonics, the amplitude of which changes rapidly in frequency domain. In contrast, the tooth's resonance response changes slowly in frequency domain and can still be perceived in the spectral envelope, when one ignores the location of the excitation harmonics, as shown in Fig.~\ref{XXX}. In Sec. XXX, we discuss how we leverage this distinction by adapting a technique that is widely used in speech processing to separate the pitch information (harmonics produced by vocal fold) and phones (spectral envelope produced by vocal tract), that is converting the signal into cepstrum domain where it can be easily separable.

% \sssec{Mitigating Alignment Errors:} While \name\ instructs users to brush specific teeth in a pre-specified order, guided by a video as they brush, it is inevitable that users make occasional errors. These errors could manifest if users have poor visibility of their own teeth (especially upper inner molars) or brush certain teeth too quickly. To limit the impact of these alignment errors on overall system performance, \name\ applies Dynamic Time Warping (DTW) to align multiple scanning sequences of the user's tooth from a same quadrant of teeth. Across multiple brushes, this allows \name\ to obtain reliable signatures of individual teeth while remaining resilient to occasional alignment errors by the user. 
\sssec{Mitigating Noise Factors and Tooth Matching Errors during Brushing:} \swarun{\name\ further addresses various sources of error that can impede its performance, such as ambient noise or user-specific brushing behavior. } \kuang{First, during the brushing procedure in real-world scenarios, the microphone on \name\ captures not only the acoustic signal we desire from the vibrating tooth, but also the sounds directly from the toothbrush, as well as various ambient noise, which introduces significant error for the signature extraction algorithm. To address this, \name\ applies Empirical Mode Decomposition~(EMD) for noise suppression to mitigate environmental noise and the direct signal from the toothbrush~(Sec.~\ref{subsec:noise-suppression}).}

\kuang{A second potential source of error occurs when users are instructed by \name\ to brush their teeth in a pre-specified order, guided by a video, to collect signatures from individual teeth. However, given the low-light conditions at some blind spots inside the mouth (such as inner upper molars), occasional matching errors can occur when a user collects measurements in these regions with poor visibility. For instance, a user may} \swarun{inadvertently} \kuang{miss brushing a tooth located in a blind spot.} \kuang{ To limit the impact of matching errors in blind spots,  \name\ developed a sequence alignment algorithm based on Dynamic Time Warping (DTW) to align brushing sequences of the user's tooth with the references from a same quadrant of teeth, which allows to obtain signatures from each corresponding tooth more accurately~(Sec.~\ref{sec:alignment}).}

We implement the prototype of \name\ using a Philips Sonicare ProtectiveClean 6100 electric toothbrush~\cite{sonicare6100} with a Voice Technologies VT500X waterproof microphone~\cite{vt500x}. We perform the experiments on dental-standard teeth models, as well as 19 users in a dental clinical center and in a home setting (IRB-approved). Our results show a health detection performance with ROC-AUC values of 0.90, 0.83, and 0.88 for detecting caries, calculus, and food impaction respectively. \kuang{An expert evaluation by interviewing two expert dentists (Sec.~\ref{sec:expert}) suggests our data collection procedure is valid and system accuracy is promising, while also pointing out both the potential benefits of integrating \name\ into the current dental healthcare system as well as potential barriers.}

\vspace*{0.05in}\noindent\textbf{\kuang{Contributions:}} \kuang{Our main contributions include:
\vspace*{-0.05in}
 \begin{itemize}
    \item The prototype of the \name\ that uses an off-the-shelf sonic toothbrush for dental health sensing.
    \item A signal processing pipeline to extract tooth resonance signatures from raw audio recordings, and perform health detection to identify dental conditions, as well as mitigate real-world noise factors.
    % \item We extract clean tooth signatures from raw toothbrush audio recordings by modeling the vibration system which enables further signal processing algorithms.
    % \item We design a tooth scanning procedure to capture signatures from a series of teeth and apply DTW-based time-series alignment to isolate the signatures from individual teeth.
    % % \item To align different measurements, we design a HMM to infer the actual location of the toothbrush, as well as use EM to learn the parameters in HMM in an unsupervised fashion.
    \item Comprehensive experiments on dental models and user studies, and an expert evaluation to validate the performance of our system.
 \end{itemize}
}
% We evaluate \name\ using a XXX toothbrush modified with our special sleeve containing a microphone. We process the collected data in the cloud using MATLAB and evaluate the performance for detecting decay and breakages typically undetected using just visual aids. Our results on dental-standard teeth models and clinical trials (under IRB) show:
% \begin{itemize}
%     \item XXX\% accuracy in isolating the behavior of the individual teeth.
%     \item YYY\% accuracy in detecting decay and breakages.
%     \item dectection of XXX decay and breakages in real patients typically undetected using just visual aids.
% \end{itemize}

% \noindent\textbf{IRB/Ethics Statement:}

% \noindent\textbf{Contributions:} Our contributions include:
%  \begin{itemize}
%     \item Developing a DIY dental diagnosis solution that uses an electric toothbrush to detect decay and breakages in human teeth.
%     \item A solution for extracting the dental jawline signature from humans and making it robust to human factors.
%     \item Evaluation of the above on dental-standard benchmarks and real world patients showcasing its efficacy.
%  \end{itemize}
\section{Background of Dental Health Problems}

\begin{figure*}
    \centering
    \begin{subfigure}[b]{0.3\linewidth}
        % \vspace*{-5in}
        % \vspace{0.05in}
        \includegraphics[width=\textwidth]{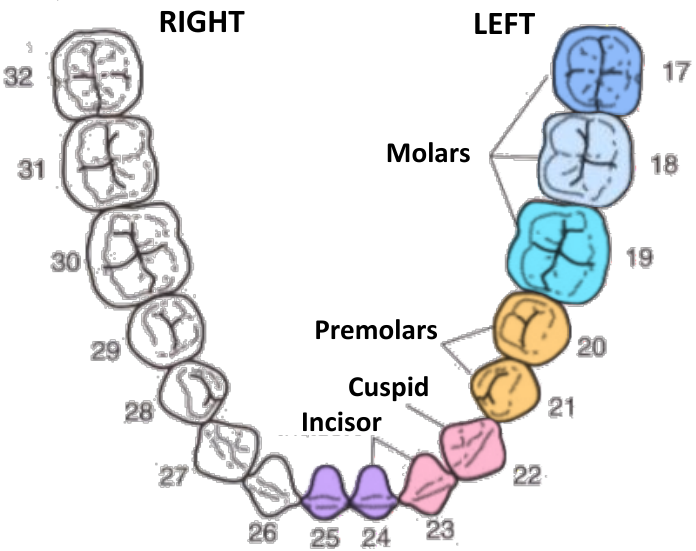}
        \caption{\rr{Chart of lower teeth of human}}
        \label{fig:tooth-chart}
    \end{subfigure}
    % \vspace{-0.1in}
    \hspace{0.2in}
    \begin{subfigure}[b]{0.58\linewidth}
        \vspace{-0.05in}
        \includegraphics[width=\textwidth]{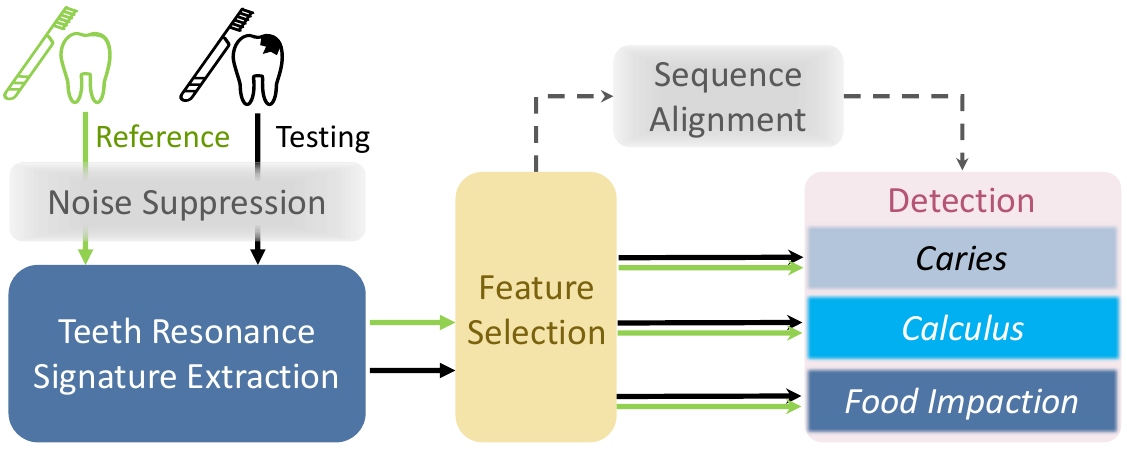}
        % \vspace{0.2in}
        \caption{\name\ detects dental diseases by comparing feature-specialized acoustic signatures with pre-collected healthy references.}
        \label{fig:overview-signalprocessing}
    \end{subfigure}
    \vspace{-0.1in}
    \caption{System overview of \name. 
    \rr{(a) showcases the brushing segments of the lower-left quadrant in different colors.} (b) shows the signal processing pipeline. }
    \vspace{-0.1in}
\end{figure*}

This paper targets the early detection of dental conditions as they may evolve over time. Theoretically, \name\ can potentially detect any dental health problem that induces significant acoustic changes to the resonant behavior of the teeth. \kuang{We have collaborated with two dental experts to identify the following three dental conditions with high prevalence, and early at-home detection of these conditions can provide valuable support in preventing more severe dental diseases.}
% This paper evaluates \name\ for detecting three of the most common dental conditions:
 
\sssec{Caries:} Tooth decay when bacteria in the mouth produce acids that demineralize the hard tissues of the teeth. Dental caries~\cite{pitts2017dental}, also called cavities, occur when teeth get permanently damaged causing pain and infection. Such damage develops into small holes that enlarge over time and may result eventually in tooth extraction. \kuang{Tooth decays have a high prevalence~\cite{preval}~(90\%) among adults, which often occur in the back teeth. Moreover, they can form between teeth~\cite{Chirichella_2022} or inside a tooth~\cite{Staff_Staff_2023}, complicating visual detection. Demineralized tissues absorb X-rays differently than hard tissues, allowing them to be identified in radiographs~\cite{nih_caries}.} This paper aims to explore a solution for at-home early detection of such decay to mitigate more serious damage to the tooth.

\sssec{Dental Calculus (Tartar):} A step toward cavities and other dental diseases is the build-up of bacteria on teeth which starts with soft deposits resulting in calculus at the end. Dental calculus~\cite{forshaw2022dental} is a hardened bacterial plaque forming below and above the gum line. \kuang{Notably, the subgingival calculus that forms beneath the gumline is not immediately visible. If left untreated, dental calculus can irritate the gums and lead to gum disease over time.}

\sssec{Interdental Food Impaction:}  Food impaction is a very common cause of gingival and periodontal disease~\cite{khairnar2013classification}. Removing the remains of food, especially those lodged between two neighboring teeth, is important to avoid the formation of plaque and eventually cavities. \kuang{Food impaction frequently happens in hard-to-see areas. \name\ can potentially serve as a tool to detect and localize the teeth with food impaction. By identifying these specific areas, brushing and cleaning can be optimized, allowing for targeted flossing at the sites of food impaction rather than relying on random flossing~\cite{sawan2022effectiveness}.}

\section{Related Work}

\textbf{Dental Condition Monitoring:} Currently, dental health monitoring is mainly available through infrequent dentist visits by using professional examination devices such as X-ray.  There has been much clinical work in dentistry using pH-based and fluorescence-based tooth-decay sensing~\cite{sharma2022ph,zakian2009near,rechmann2014soprocare,pretty2005quantification, hibst2001detection,pretty2006caries, wong2017dental,angelino2017clinical}. \kuang{Some recent work has explored incorporating sensing technologies into the regular toothbrushing procedure. However, most of these related systems focus on monitoring hygiene behaviors by detecting which part of the tooth a patient is currently brushing, such as using magnetic field sensing~\cite{MET}, sound being produced~\cite{AsymSound, ToothbrushSound}, camera~\cite{learntoothbrush}, and wrist-worn motion sensors~\cite{mTeeth, mOral, luo2018brush, huang2016}. 
% Similarly, another line of work~\cite{mTeeth, mOral, luo2018brush, huang2016} uses wrist-worn motion sensors for brushing behavior tracking. 
Besides, LumiO~\cite{LumiO} integrates a blue-violet light intraoral camera into the toothbrush head to capture the intraoral images and monitor the progress of plaque removal, which does not focus on dental disease detection.} More recently, OralCam~\cite{OralCam} has proposed to use a smartphone camera for visible dental disease detection. \kuang{However, OralCam can only detect visually diagnosable diseases and struggles in low-light conditions inside the mouth. To conclude, dental health monitoring using ubiquitous devices is an underexplored area, and \name\ has the unique advantage of identifying invisible diseases such as inner cavity compared with existing camera-based solutions.}

\sssec{Acoustic Sensing on Humans:} Acoustic sensing has been widely explored recently for gesture recognition~\cite{cao2020earphonetrack, nandakumar2016fingerio, sun2018vskin} as well as health monitoring~\cite{headFi, zhang2020your,wang2018unlock, song2020spirosonic, respiratory, wang2018c, xu2019breathlistener, apnea, opioid}. Specifically, acoustic signals have been utilized for monitoring heart rate~\cite{headFi, zhang2020your,wang2018unlock}, lung function\cite{song2020spirosonic}, sleep apnea~\cite{apnea}, and respiratory rate~\cite{respiratory, wang2018c, xu2019breathlistener}. The majority of this line of work counts on detecting movements of the human skin surface for monitoring health conditions. On the other hand, \name\ leverages the tooth resonance behavior caused by the electric toothbrush, for extracting unique signatures for monitoring the health status of the teeth.

\sssec{Sensing with Acoustic Resonance:} Measuring the acoustic resonance of an object can reflect its integrity and structure. Therefore, past work has explored acoustic vibrometry for detecting defects including cracks in cups~\cite{milton}, gear tooth breakage~\cite{singh1998detecting}, railroad defects~\cite{gong1992acoustic}, and fruit ripeness~\cite{galili1998acoustic,schotte1999acoustic}.  While this prior work has explored a wide set of applications using acoustic vibrometry, it detects mainly big objects and uses a dedicated transmitter as the excitation source. Compared to prior work, \name\ uses an off-the-shelf sonic toothbrush as the excitation source, which is not precisely controllable, and can detect the defects on human teeth.

%%%% search for RF/mmWave health/dental sensing
\section{overview}
\label{sec:overview}

\name\ is a dental health sensing system based on an off-the-shelf sonic toothbrush. \name\ begins by collecting a set of reference signatures from the user's teeth when they brush for the first time with the system. Note that we require that new users visit a dentist and assume all teeth are in a relatively healthy status post-visit. After collecting the reference signatures, whenever the user wants to perform a self-examination, they brush their teeth to collect the signature again for testing. 
% \name\ instructs the user to move their toothbrush over each tooth through a video-guided pattern that takes less than two minutes overall. The video guides the user to brush specific quadrants of the teeth at each time, moving slowly over the individual teeth to ensure coverage of all teeth. 
\name\ compares this test signature with the reference signatures, and notifies the user if it detects potential dental problems.
We further use the signature from different measurements across days to boost the detection accuracy, which enables a more accurate understanding of one's dental health status through long-term monitoring.

\sssec{Data Collection:} \kuang{\name\ instructs the user to perform the measurement through a video-guided pattern. The video guides the user to brush their teeth through the order of four teeth quadrants: upper right, upper left, lower right, and lower left. For the brushing in each quadrant, the video guides the user to brush the teeth one or two at a time, from the innermost molar to the outermost incisor, and stay on the chewing surface of each tooth for a few seconds. Fig.~\ref{fig:tooth-chart} shows the structure of the lower teeth, and the colors illustrate the brushing segments for the lower-left quadrant as an example. The user starts brushing from three molars (from 17 to 19) and brushes one at a time. For the premolars, cuspid, and incisors, as the size of the tooth is generally smaller than the regular toothbrush head, the video guides the user to brush two of the teeth together at a time (20\&21, 22\&23, and 24\&25). \name\ captures an audio clip for each single tooth/two teeth brushing and performs signal processing to extract the individual tooth signature. We also use the whole scanning sequence of each tooth quadrant to perform sequence alignment to mitigate the matching errors.}

\sssec{System Overview:}
Fig.~\ref{fig:overview-signalprocessing} presents the processing pipeline of \name. \swarun{\name's primary contribution is it's health detection pipeline (Sec.~\ref{sec:signature}-\ref{sec:detection}) that consists of three key steps: (1) \textit{Tooth Resonance Signature Extraction: } We slice the audio recording from the microphone into chunks of 50 milliseconds, for each of which we extract a resonance signature of the tooth that was vibrated over that specific time span. (2) \textit{Feature Selection: } We then process this signature to further extract unique features that assist in identifying different dental conditions. (3) \textit{Health Detection: } To detect each of the diseases, we aggregate and average the features from the same tooth \kuang{among the 50-millisecond signal chunks}, and finally compare them with the reference features captured from the corresponding tooth. }

\swarun{We further note two important processing steps for improving system resilience to errors and noise (Sec.~\ref{sec:factors}). (1) \textit{Noise suppression: } For every raw audio recording captured by the microphone on \name, we first perform noise suppression to make the signal focused on the vibration behavior of the teeth and toothbrush. (2) \textit{Sequence Alignment: } Before aggregating the features and performing health detection, \name\ uses the features from each chunk to perform a time-series alignment algorithm to find the best match of each signature to the corresponding tooth, which mitigates the matching errors for brushing in blind spots.} 

\kuang{In the following sections, we first discuss how \name\ model the vibration behaviors during toothbrushing and extract tooth resonances signature~(Sec.~\ref{sec:signature}), then elaborate how we apply the feature selection~(Sec.~\ref{sec:feature}) and health detection~(Sec.~\ref{sec:detection}) to detect different dental conditions. We also detail the noise suppression sequence and alignment algorithms in Sec.~\ref{sec:factors}.}

\section{Tooth Signature Extraction}
\label{sec:signature}

\kuang{In this section, we discuss how we extract tooth resonance signatures from raw audio recordings captured by the microphone during toothbrushing. \name\ first applies Short-Term Fourier Transform~(STFT) with 50 milliseconds as time window. Following this, we explain how our signature extraction extracts the resonance signature from the tooth (or two teeth) brushed upon over a single 50-millisecond time window.}

% In this section, we discuss how we extract tooth resonance signatures from raw audio recordings captured by the microphone during toothbrushing. We assume that every such signature is obtained over a 50 millisecond time-window during which one or more teeth are brushed upon (usually one tooth if the brush were exactly above it or two if at the boundary of two teeth). For simplicity, we consider the case of exactly one tooth being brushed upon in this section over the 50 millisecond duration. We note that in practice, the same tooth could be brushed upon for much longer than 50 milliseconds. We will deal with recognizing and resolving such situations in Sec.~\ref{sec:alignment}. For now, let us first model the vibration of an individual tooth in Sec. \ref{subsec:vibration-model} below, that uses the Short-Term Fourier Transform~(STFT) over a 50 millisecond time-span of the audio sequence received from the microphone. Following this, we explain our noise suppression and signature extraction algorithms.

\subsection{Vibration Behavior Modeling}
\label{subsec:vibration-model}

\begin{figure*}[t]
\begin{minipage}{0.58\textwidth}
\centering
  \begin{subfigure}[b]{0.5\linewidth}
    \includegraphics[width=\textwidth]{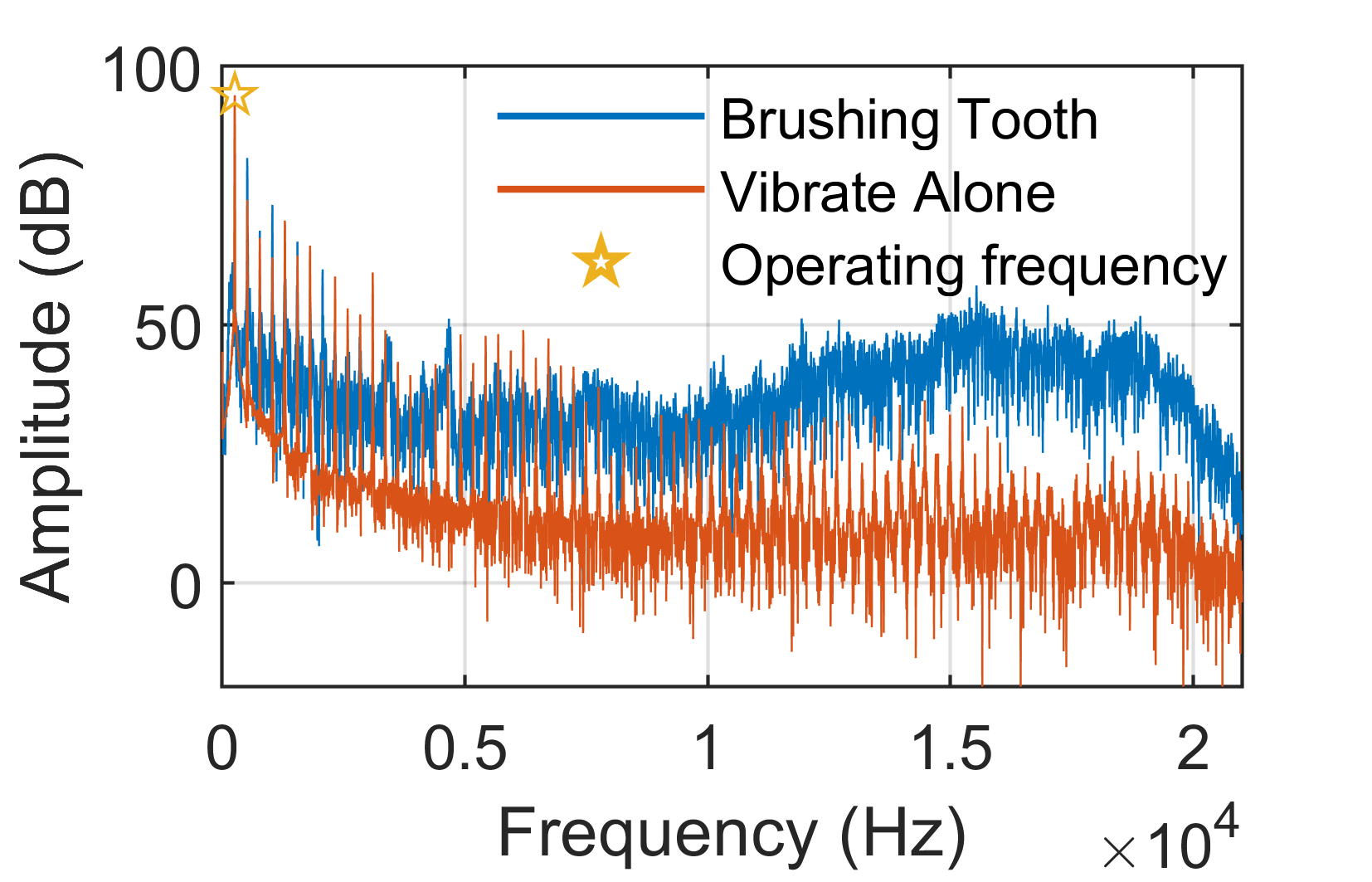}
    \caption{Microphone signal}
    \label{fig:sig-mic}
  \end{subfigure}
  \hspace{-0.12in}
  \begin{subfigure}[b]{0.5\linewidth}
    \includegraphics[width=\textwidth]{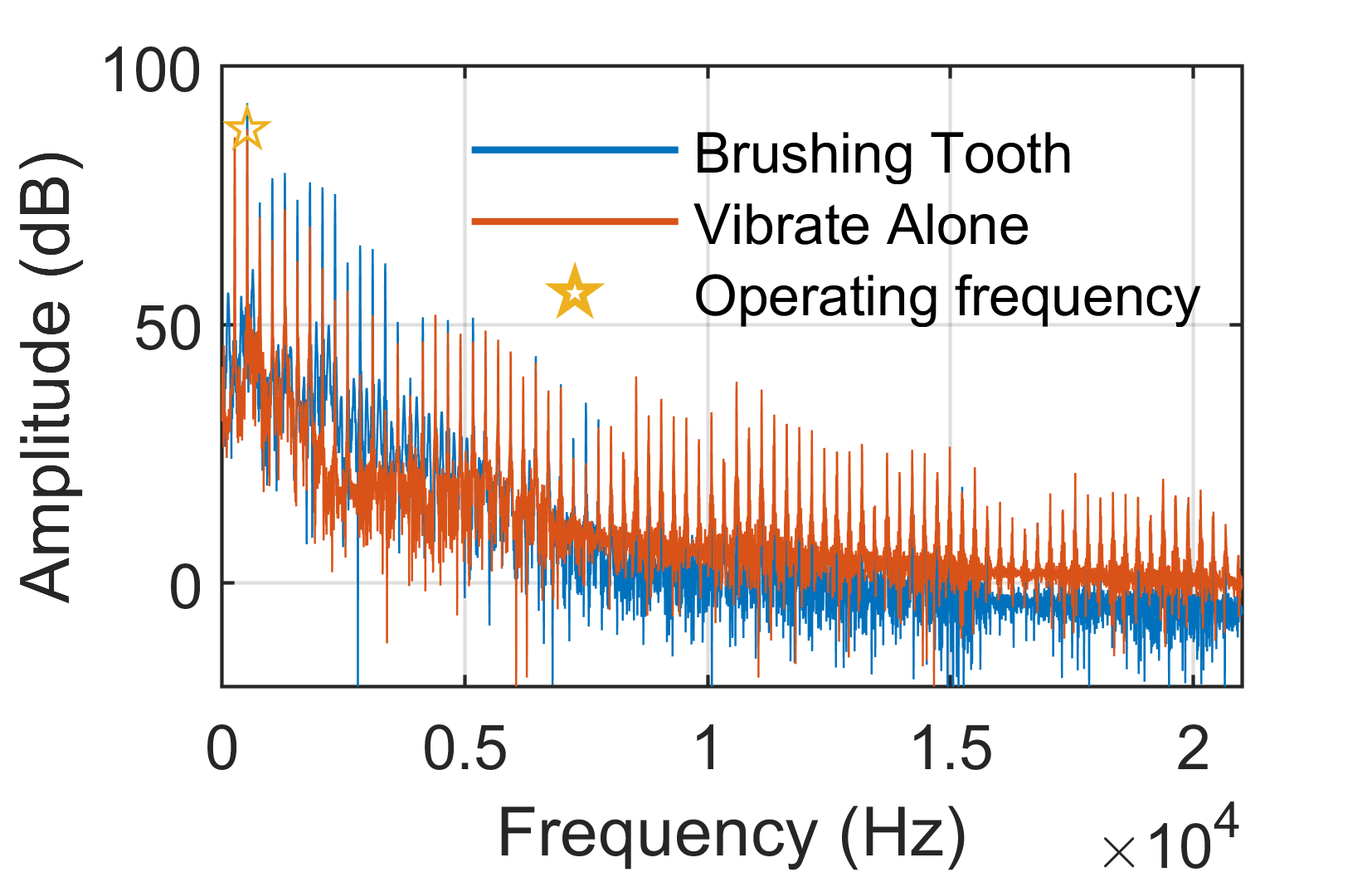}
    \caption{Vibration sensor signal}
    \label{fig:sig-vib}
  \end{subfigure}
  \vspace{-0.1in}
  \caption{Signal spectra when a sonic toothbrush is vibrating alone vs. when in contact with a tooth. (a) shows brushing induces the resonances of the tooth. (b) shows brushing also changes the excitation signal itself.}
  \label{fig:figure1}
\end{minipage}
\hspace{0.1in}
\begin{minipage}{0.38\textwidth}
    \centering
     \includegraphics[width=\linewidth]{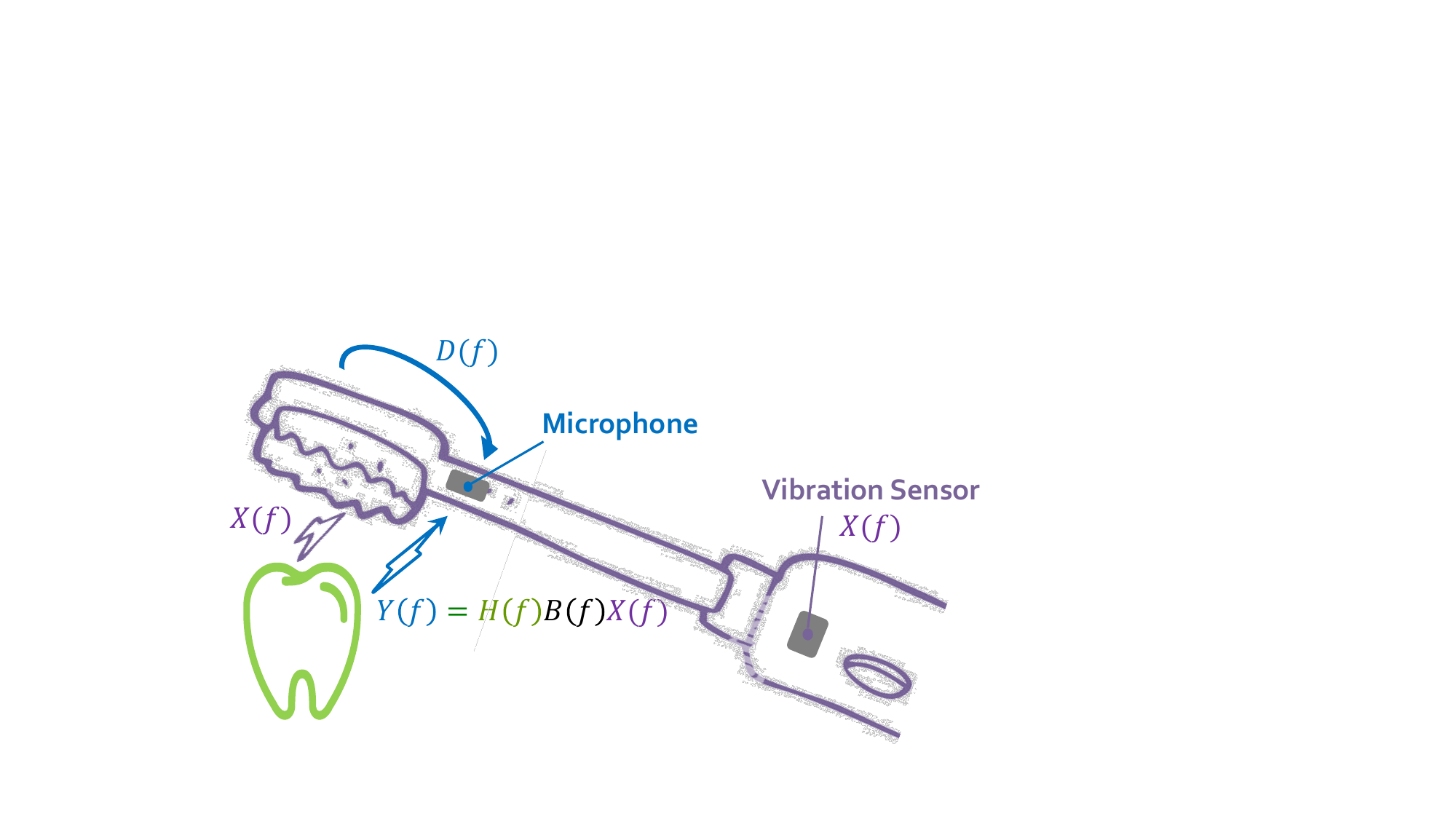}
     \vspace{-0.15in}
    \caption{The signal received by the microphone composed of the excitation signal directly from the toothbrush head as well as the tooth response}
    \label{fig:paths}
\end{minipage}
\vspace{-0.1in}
\end{figure*}

Sonic toothbrushes are a subset of electric toothbrushes with movement that is fast enough to produce vibration in the audible range. Typically, an off-the-shelf sonic toothbrush is designed to vibrate at a specific frequency in the range of 200-400~Hz. It also produces harmonics in higher frequencies which are integer multiples of the operating fundamental frequency. The red line in Fig.~\ref{fig:sig-mic} shows the spectrum of the acoustic signal produced by the Philips Sonicare ProtectiveClean 6100 when it is turned on. While it is designed to operate at 260~Hz (see the yellow star), it also generates resonant tones at higher frequencies up to around 20~kHz. 

The core idea of \name\ leverages this special characteristic of the sonic toothbrush and uses it as a wideband excitation source for acoustic vibrometry for sensing the teeth. In a typical vibrometry system, the user applies excitation signal from sources such as piezoelectric transducers to induce vibrations in the object being studied~\cite{milton,clark2004rail,industryobjectcrackanalysis}. The resonance behavior of the object can then be deduced and measured. Mathematically, we can describe the acoustic behavior of a simple probe-based system as~\cite{enwiki:frequency-response}:
\begin{equation}
    Y(f) = H(f)X(f) \label{eqn:y}
\end{equation}
where $X(f)$ is the excitation signal, $H(f)$ is the resonance signature of the object, and $Y(f)$ is the response signal being captured.  Generally, one has a firm knowledge of $X(f)$ produced by a specialized source, so by simply deconvolving $X(f)$ from $Y(f)$ (division in frequency domain), the signature $H(f)$ can be obtained.

When a sonic toothbrush is in contact with a tooth, it also shows a similar resonance phenomenon. As shown in Fig.~\ref{fig:sig-mic}, when the toothbrush is brushing a tooth, the resonant behavior of the tooth is induced, making the amplitude of the signal amplified at that frequency~(2-20~kHz) compared to vibrating in isolation. However, compared to the typical vibrometry system, \name\ faces three unique challenges when we want to extract the resonance signatures:
\begin{enumerate}
    \item The excitation signal especially the harmonic peaks in higher frequencies are constantly changing since they are essentially noise produced by the imperfect vibration motor.\label{vib-challenge1}
    \item The excitation signal would be altered due to the contact with teeth. To confirm this, we attach a vibration sensor to the toothbrush body \kuang{(illustrated in Fig.~\ref{fig:paths})} to capture its source signal, the measurement of which is shown in Fig.~\ref{fig:sig-vib}. We can clearly see the trend of amplitude changing across the spectrum when it is brushing (Blue line) versus vibrating alone (Red line). \label{vib-challenge2}
    \item The toothbrush head and the tooth are not in perfect contact because of the soft bristles. As a result, different brushing strengths, tiny movements and slippages can produce slightly different acoustic responses from the teeth. \label{vib-challenge3}
\end{enumerate}

To formulate (1), we denote the excitation signal emitted by the motor ${X}(f)$, which is not a constant signal, but instead a harmonic source with constantly changing harmonics. To model (2) and (3) -- the impact of bristle contact with the tooth on signal, we denote by $B(f)$, the overall change in the response signal this causes. These two terms in consonance allow us to re-write Eq.~\ref{eqn:y} as: 
\begin{equation}
% \vspace{-0.1in}
\begin{split}
\label{eq:ywithb}
    Y(f) = H(f)B(f){X}(f) 
\end{split}
% \vspace{-0.1in}
\end{equation}

%To formulate (\ref{vib-challenge1}) and (\ref{vib-challenge2}), we can model the excitation signal as:

% \begin{equation}
%     X(f) = B_1(f)\widetilde{X}(f)
% \end{equation}
% where $\widetilde{X}(f)$ represent the constantly changing harmonics source, and $B_1(f)$ is the excitation signal changes induced by the contact \swarun{The difference between $X$ and $\tilde{X}$ is not clear}.

% Further, considering (\ref{vib-challenge3}), the response signal of the tooth being vibrated can be formulated as:
% % \begin{equation}
% % \begin{split}
% %     Y(f) = H(f)B'(f)X(f) &= H(f)B'(f)B_1(f)\widetilde{X}(f) \\
% %                         &= H(f)B_2(f)\widetilde{X}(f)
% % \end{split}
% % \end{equation}
% % where $B'(f)$ is the response changes caused by different brushing strengths, movements and slippages of the toothbrush head. $B_2(f)$ is the overall changes to the response signal caused by the contact.

% \begin{equation}
% \begin{split}
%     Y(f) = H(f)B_2(f)\widetilde{X}(f) 
% \end{split}
% \end{equation}
% where $B_2(f)$ is the overall changes to the response signal caused by the contact, which originated from both (\ref{vib-challenge2}) and (\ref{vib-challenge3}).

Next, note that our microphone is not in direct contact with the teeth. Instead, it receives signals traveling through the air from both the teeth and the toothbrush. This means that in addition to $Y(f)$, the microphone also hears signals from the direct path $D(f)$ between the toothbrush and microphone, \kuang{as well as other environmental noise $N(f)$. As shown in Fig.~\ref{fig:paths}, the signal received by the microphone can be denoted as:}
\begin{equation}
% \vspace{-0.1in}
% \begin{split}
\label{eq:whole-model}
    M(f) = Y(f) + D(f) + N(f) \\
    = H(f)B(f){X}(f) + D(f) + N(f)
% \end{split}
% \vspace{-0.1in}
\end{equation}

% \kuang{Is here clear enough? I'm thinking of adding a figure to illustrate these two paths.}

\kuang{So far, we have modeled the vibration behavior of the system including toothbrush excitation ${X}(f)$, resonance behavior of the tooth $H(f)$, other effects caused by the contact $B(f)$, as well as direct path signal $D(f)$ and ambient noise $N(f)$.  We will elaborate in Sec.~\ref{subsec:noise-suppression} how we perform noise suppression based on EMD to mitigate the effect $D(f)$ and $N(f)$. For now, we assume that the signal we used to proceed with the signature extraction algorithm can be approximately represented by Eq.~\ref{eq:ywithb}.}
% of and Sec.~\ref{subsec:extract-alg} how we can develop noise suppression and signature extraction algorithms to estimate the teeth signatures $H(f)$ from this complex response.

% Thus far, the excitation signals directly coming from the toothbrush are mostly canceled, especially for the part in the frequency range under 3~kHz

\subsection{Extraction Algorithm}
\label{subsec:extract-alg}
%\kuang{Before further processing, note that the audio signal we captured is a time-series sequence that includes a series of teeth signatures. Thus, we first perform Short-Time Fourier Transform~(STFT) on the denoised audio. We assume that the signature inside a small STFT time window is stationary and only contains signatures from a single tooth or a fixed combination of adjacent teeth.}

\kuang{Given Eq.~\ref{eq:ywithb}, our goal is to extract the teeth resonance signature $H(f)$ from the signal $Y(f)$.} However, we also have two undesirable terms:  ${X}(f)$, the constantly changing harmonics, and $B(f)$, related to strength, movement and other artifacts during the process of brushing. It is infeasible to measure either of these in real-time to divide them out.

To tackle this challenge, we make a key empirical observation: while these three terms share the same frequency band, their rates of change across the frequencies are different.

% To better understand this, consider the logarithmic spectrum of a signal as shown in Fig.~\ref{fig:envelope}. We can clearly see that it has a periodic structure of a series of spikes, which correspond to the harmonics in the excitation signal -- unsurprising, given the toothbrush emits a series of harmonic tones. These harmonics may be attenuated or strengthened at different frequencies based on the strength of bristle contact. Importantly, we see that Fig.~\ref{fig:envelope} also has a macro-level structure, besides these harmonics, that owes its origin to the response of the tooth itself. We can better observe this macro-structure by connecting the valleys of the harmonic structure. This is called the spectral envelope, which is composed of slow-changing peaks and valleys (shown in red in  Fig.~\ref{fig:envelope}). This envelope corresponds to the tooth's resonance and is free of the acoustic emissions of the toothbrush or indeed the impact of the bristles. 

\begin{figure}[h]
\centering
\vspace{-0.15in}
  \begin{subfigure}[b]{0.5\linewidth}
    \includegraphics[width=\textwidth]{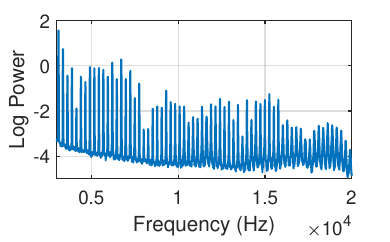}
    \vspace{-0.2in}
    \caption{Vibrating alone}
    \label{fig:envelope-notouch}
  \end{subfigure}
  \hspace{-0.12in}
  \begin{subfigure}[b]{0.5\linewidth}
    \includegraphics[width=\textwidth]{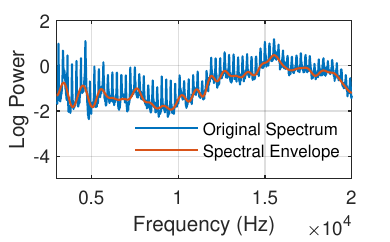}
    \vspace{-0.2in}
    \caption{Brushing}
    \label{fig:envelope-touch}
  \end{subfigure}
  \vspace{-0.1in}
  \caption{Examples of signal in log spectrum (denoised version) when the toothbrush is vibrating alone vs. brushing}
  \label{fig:envelope}
\vspace{-0.1in}
\end{figure}

\begin{figure*}[t]
  \centering
  \begin{subfigure}[b]{0.21\linewidth}
    \includegraphics[width=\textwidth]{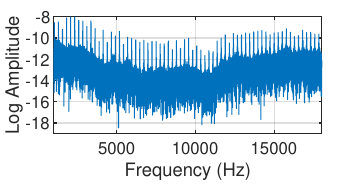}
    \caption{Raw Log Spectrum}
    \label{fig:cep-1-raw}
  \end{subfigure}
  \hspace{-0.15in}
  \begin{subfigure}[b]{0.21\linewidth}
    \includegraphics[width=\textwidth]{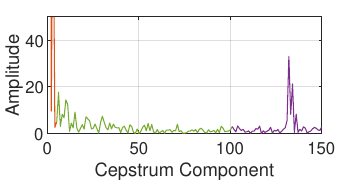}
    \caption{Cepstrum}
    \label{fig:cep-1-cep}
  \end{subfigure}
  \hspace{-0.15in}
  \begin{subfigure}[b]{0.21\linewidth}
    \includegraphics[width=\textwidth]{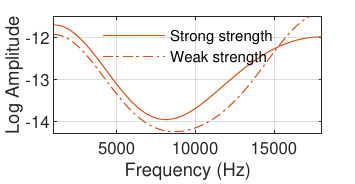}
    \caption{Lower Components}
    \label{fig:cep-1-low}
  \end{subfigure}
  \hspace{-0.15in}
  \begin{subfigure}[b]{0.21\linewidth}
    \includegraphics[width=\textwidth]{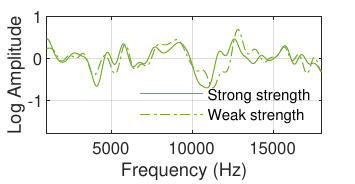}
    \caption{Signature Components}
    \label{fig:cep-1-mid}
  \end{subfigure}
  \hspace{-0.1in}
  \begin{subfigure}[b]{0.21\linewidth}
    \includegraphics[width=\textwidth]{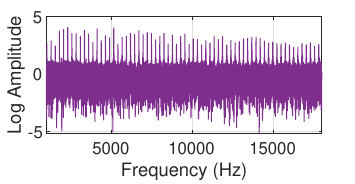}
    \caption{Higher Components}
    \label{fig:cep-1-high}
  \end{subfigure}
  \hspace{-0.2in}
  \vspace{-0.1in}
  \caption{Signature Extraction algorithm by transforming the signal into cepstrum domain. The lower red components correspond to brushing artifacts such as strength and movement. The green components in the middle are the desired teeth resonance signatures. The higher purple components correspond to the excitation harmonics.}
  % \vspace{-0.1in}
  \label{fig:cep-1}
\end{figure*}

To better understand this, consider the logarithmic spectrum of a signal shown in Fig.~\ref{fig:envelope}. We can clearly see that both the signal when vibrating alone~(Fig.~\ref{fig:envelope-notouch}) and the signal during brushing~(Fig.~\ref{fig:envelope-touch}), have the periodic structure of a series of spikes, which correspond to the harmonics in the excitation signal. Importantly, we see that the signal produced during brushing (Fig.~\ref{fig:envelope-touch}) also has a macro-level structure, besides these harmonics, that owes its origin to the tooth's response. We can better observe this macro-structure by connecting the valleys of the harmonic structure. This is called the spectral envelope, which is composed of slow-changing peaks and valleys (shown in red in  Fig.~\ref{fig:envelope-touch}). By comparing Fig.~\ref{fig:envelope-notouch} and Fig.~\ref{fig:envelope-touch}, we observe that the envelope corresponds to the tooth's resonance and is free of the excitations from the toothbrush.

% \begin{figure}[h]
%     \centering
%     \vspace{-0.1in}
%     \includegraphics[width=0.8\linewidth]{figures/envelope.pdf}
%     \vspace{-0.1in}
%     \caption{An example of an denoised acoustic measurement in log spectrum domain where the spectral envelope captures the resonant behavior of the tooth}
%     \vspace{-0.1in}
%     \label{fig:envelope}
% \end{figure}

One way of evaluating and separating these periodic structures in a signal changing at different scales/speeds is to use the Fourier Transform. Specifically, we take the Discrete Cosine Transform~(DCT) of the log spectrum, to obtain the \textit{cepstrum}\footnote{a word-play on spectrum, commonly used in human speech analysis~\cite{cepstrum-wiki}} representation of the signal. This domain disentangles the three components described above cleanly across frequencies.%The name is a word-play as it essentially backward transforms the signal in the \textit{spectrum}.

Mathematically, we transform the $M_p(f)$ into cepstrum domain as follows:
\begin{equation*}
\begin{split}
    \mathcal{D}(\log|M_p(f)|) \approx \mathcal{D}(\log|{X}(f)|+\log|H(f)|+\log|B(f)|) \\
                            = \mathcal{D}(\log|{X}(f)|)+ \mathcal{D}(\log|H(f)|) + \mathcal{D}(\log|B(f)|)
\end{split}
\end{equation*}
where $\mathcal{D}$ represents the DCT. In this way, we convert the multiplicative components in the spectrum domain into additive components in the cepstrum domain where they are separable. The $\mathcal{D}(\log|H(f)|)$ term is the desired tooth resonance signature.

Fig.~\ref{fig:cep-1-cep} shows an example of converting the signal in Fig.~\ref{fig:cep-1-raw} into the cepstrum domain. Through detailed empirical analysis, we find that the three segments in different colors contain three different parts of information. First, we can clearly see a peak in higher cepstrum components colored in {\color{violet} purple} that represent the harmonic structure of the excitation signal ${X}(f)$, as it is composed of rapidly-changing spikes in the log spectrum. The location of the peak corresponds to the fundamental frequency of the harmonics. Next, the components at the middle cepstrum frequencies colored in {\color{green}green} represent the tooth resonance signatures $H(f)$. We confirm this empirically, by observing that this region of the spectrum remains consistent when the same tooth is brushed multiple times, even with different strengths or slippages. Indeed, most brushing artifacts such as strength and movements mainly influence the few lowest cepstrum components colored in {\color{red}red}, representing $B(f)$. \cref{fig:cep-1-low,fig:cep-1-mid,fig:cep-1-high} show the results of converting these three components back to the log spectrum domain. We can clearly see they represent the structure on different scales of changes in the spectrum and have measure the three different physical behaviors. 

One might wonder if the extracted $H(f)$ (the green plot in Fig.~\ref{fig:cep-1-mid}) is robust to differences in brushing behavior. To evaluate this, we illustrate a trace of the extracted components from the measurements on the same tooth but with different brushing strengths (we present a more robust multi-user evaluation later in results in Sec.\ref{sec:new-evaluation} and this is only a representative example to illustrate the point). In Fig.~\ref{fig:cep-1-low} and Fig.~\ref{fig:cep-1-mid}, the dashed lines denote brushing with weak strength and the solid line for strong strength. We can clearly see the lower components in Fig.~\ref{fig:cep-1-low} have significant changes between the two brushing strengths, while the signatures in Fig.~\ref{fig:cep-1-mid} are still very close to each other, especially when observing the location and amplitude of each peak and valley. This decomposition result showcases that the signature we extract is robust to differences in brushing behavior, and is robust to constantly changing excitation harmonics.

% We further show the signal with higher brushing strength for the same tooth in Fig.~\ref{fig:cep-2}. We can clearly see that the amplitudes corresponding to the lower cepstrum components in Fig.~\ref{fig:cep-2-low} have significant changes compared with Fig.~\ref{fig:cep-1-low}, while the teeth signatures we extracted shown in Fig.~\ref{fig:cep-2-mid} is still very close to that in Fig.~\ref{fig:cep-1-mid}, especially the location and amplitude of each peak and valley. 

\section{Feature Selection}
\label{sec:feature}

The goal of \name\ is to compare the tooth signature from any individual tooth being brushed with the corresponding healthy references, and compute a probability of each disease of interest for every single tooth. \name\ seeks build a platform that can potentially detect a wide range of dental diseases. Different diseases are likely to induce changes to different parts of the signature. Using the signature as a whole from Sec.~\ref{sec:signature} may not be effective enough to distinguish different kinds of dental problems. One therefore desires more fine-grained signatures for different specific tasks (e.g. caries detection), to distinguish between healthy and unhealthy teeth.

In the data science community, researchers have developed contrastive learning frameworks based on Deep Neural Networks~(DNN) to distinguish between similar or dissimilar pairs of data points in other contexts. However, DNN training of contrastive learning needs at least weekly labeled data~\cite{schroff2015facenet} or data augmentations~\cite{chen2020big} from a relatively large dataset. Collecting a large real-world dataset using our new \name\ platform is out of the scope of this work.

% \name\ selects more fine-grained features among the signature for every different disease of interest. The core idea of feature selection is to identify a sub-space of signature for each disease of interest where the separability of the healthy and unhealthy tooth is maximized, thereby providing fine granularity information as well as more resilience to noise for the classification task. 

% Instead, as the first system seeking to analyze the acoustic resonance of teeth for dental health sensing, we choose the most straightforward approach -- to select a part of features among the extracted teeth resonance signature, which can best distinguish the location of the tooth brushed. Specifically, we choose these features by maximizing the separability between different teeth. In other words, for the task of identifying if two teeth brushed across different time chunks are same or different, we seek to separate the teeth at different locations on a single user.

Instead, as the first system seeking to analyze the acoustic resonance of teeth for dental health sensing, we choose the most straightforward approach -- to select a part of features among the extracted teeth resonance signature, which can best characterize if a specific disease occurs. The core idea of feature selection is to identify a sub-space of signature for each disease of interest where the separability of the healthy and unhealthy tooth is maximized, thereby providing fine granularity information as well as more resilience to noise and other factors for the classification task. 

We modified the classical discriminant analysis~(LDA~\cite{balakrishnama1998linear}), a statistical algorithm modified to our specific feature selection problem. From Sec.~\ref{sec:signature}, we extract tooth signature $\mathcal{D}(\log|H(f)|)$ represented by a series of DCT components (the points in the green plot of Fig.~\ref{fig:cep-1-cep}), denoted as $(h_1, h_2, ..., h_s)$, where $s$ is the length of a signature. Mathematically, we quantify the separability of each feature in the signature using the ratio of the between-class variance and the in-class variance. We define the gain for feature $h_i$:
\vspace{-0.1in}
\begin{equation}
G(h_i) = \frac{S_{b}(h_i)}{S_{w}(h_i)}
= \frac{\sum\limits_{k=1}^{K}N_k(\Bar{h}_{i,k}-\Bar{h}_i)^2}{\sum\limits_{k=1}^{K}\sum\limits_{n\in \mathcal{C}_k}(h_{i}[n]-\bar{h}_{i,k})^2}
% \vspace{-0.05in}
\end{equation}
where $S_{b}(h_i)$ is the between-class variance that quantifies how far the collected data samples from different classes (teeth locations) stray from each other on feature $h_i$. $S_{w}(h_i)$ is the within-class variance that quantifies how compact the samples from the same class are on feature $h_i$. $K$ is the number of classes (two classes: healthy/unhealthy in this case). $\mathcal{C}_k$ corresponds to the set of data samples in class $k$ and $N_k$ is the number of samples in class $k$. $\bar{h}_{i,k}$ is the mean value of feature $i$ in class $k$, and $\bar{h}_{i}$ is the mean value of all samples.

\kuang{Note that any feature $h_i$, offers different gains for different types of diseases based on the statistical distribution of our collected data. Thus, for every type of disease,} we choose the set of features that provide the highest gain by maximizing:
\begin{equation}
    \hat{M}, \hat{N} = \max_{M,N} \sum_{i=M}^{N} (G(h_i) - \alpha)
\end{equation}
where $\alpha>0$ is a tunable parameter to control the number of selected features. Instead of selecting individual features, we select a continuous range of DCT components in the cepstrum domain denoted as $(h_{\hat{M}}, h_{\hat{M}+1}, ..., h_{\hat{N}})$ with a starting index $\hat{M}$ and ending index $\hat{N}$,  to preserve the physical meaning of the representation. \kuang{We select this set of features as they best characterize disease occurrence on teeth. We also note that the feature selection is only performed on a partial validation dataset.}

\section{Health Detection}\label{sec:detection}

\begin{figure}
\centering
    % \vspace{0.05in}
    % \hspace{0.5in}
     \includegraphics[width=0.87\linewidth]{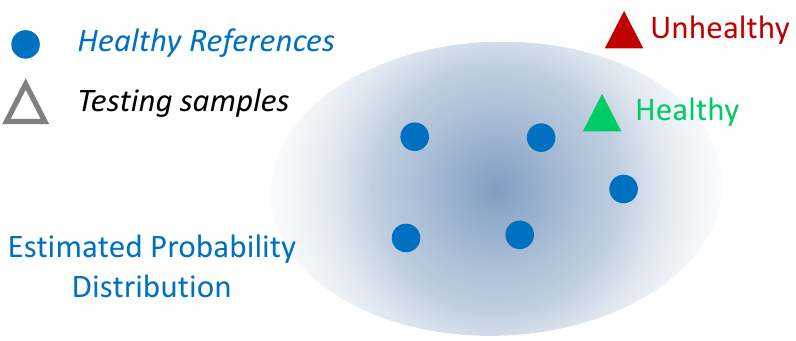}
     % \hspace{-0.5in}
     \vspace{-0.1in}
    \caption{Health detection by estimating the distribution of healthy references and thresholding the likelihood of the testing samples.}
    \vspace{-0.1in}
    \label{fig:detection}
\end{figure}

% The goal of \name\ is to compare the tooth signature from any individual tooth being brushed with the corresponding healthy references, and compute a probability of every kind of disease of interest for every single tooth.

% Note that \name\ seeks to detect a wide range of dental problems such as caries and calculus. Different diseases are likely to induce changes to different parts of the signature. Using the extracted signature from Sec.~\ref{sec:signature} as a whole may not provide the granular information needed to distinguish different kinds of dental problems. Thus, before performing the health detection tasks, \name\ selects more fine-grained features using the technique introduced in Sec.~\ref{sec:feature} for every different disease of interest. For each kind of disease, there are only two classes -- healthy and unhealthy. The core idea of feature selection is to identify a sub-space of signature where the separability of the healthy and unhealthy tooth is maximized, thereby providing more resilience to noise for the classification task. 
\kuang{After selecting features specialized for each kind of disease in every 50-millisecond time window, we aggregate and average the features from the same tooth among the time windows to improve the signal-to-noise ratio of the features. \name\ then uses the aggregated features to perform health detection for diseases of interest.}
% With the selected features specialized for each kind of disease, we then perform health detection for disease of interest. 
The most straightforward idea one may have is to use standard classification techniques such as SVM~\cite{cortes1995support} to separate healthy and unhealthy teeth. However, since the tooth signature distributions of different teeth are very different, one would require training a classifier for every individual tooth from every different user. This of course is impractical because we will not have any unhealthy samples to train the classifier before the disease occurs!

Instead, given the healthy reference tooth signatures,  \name\ designs a one-class classifier for health detection using only healthy samples. Specifically, \name\ estimates the probability density function of the healthy tooth signatures using the references, which essentially builds a profile for the original healthy tooth. The idea is visualized in Fig.~\ref{fig:detection}. When a new measurement (testing sample) from the same tooth is recorded, we compute its likelihood given the estimated distribution. We consider the tooth as an unhealthy tooth if this computed likelihood is lower than a threshold. Note that every different disease on every single tooth requires a separate one-class classifier.

% Our solution is given a different set of features extracted for each disease, we design an anomaly detector using a 1-class classifier. Our assumption is that the user starts using our toothbrush on healthy teeth, and \name gradually builds a profile for each healthy tooth. Using this profile, we can detect if a tooth has developed a disease or not, by essentially comparing current tooth signature with the history profile of the same tooth. 

Mathematically, as the size of the reference samples is small and the underlying distribution is unknown, we model the probability distribution using Kernel Density Estimation~(KDE)~\cite{hastie2009elements}. The probability density function can be estimated by:
% \vspace{-0.1in}
\begin{equation*}
    \hat{f}_h(x)=\frac{1}{n h} \sum^n_{i=1} K(\frac{x-x_i}{h})
\end{equation*}
where $X=(x_1,x_2,..,x_n)$ are the collected features from $n$ healthy references. Note that we also perform normalization for each dimension of the feature before estimating the model. $K$ is the kernel function that we empirically choose to be Gaussian. $h$ is a smoothing parameter to control the variance of the Gaussian kernel.
% \textbf{Gaussian Model of Healthy Tooth.} As the size of the data and the intrinsic dimension is relatively small for each tooth~\cite{kuzborskij2016naive}, we model each healthy tooth using Kernel Density Estimator (KDE)~\cite{hastie2009elements}:
% \begin{equation*}
%     \hat{f}_h(x)=\frac{1}{n h} \sum^n_{i=1} K(\frac{x-x_i}{h})
% \end{equation*}
% where we collect $X=[x_1,x_2,..,x_n]$, independent and identically distributed samples drawn from a multivariate distribution with an unknown density $ƒ$, and $h$, known as bandwidth, is a smoothing parameter. We use a Gaussian basis functions for the kernel function $h$.

To enable health detection, we evaluate the similarity of a new coming measurement with the healthy references using log-likelihood. Moreover, since the residual environmental noise can impact the tooth signature which may further produce false positive results, we bootstrap the detection accuracy using the data across different measurements and even across days. Specifically, we assume signatures across different measurements are independent. We can compute the aggregated log-likelihood:
% \vspace{-0.1in}
$$
\log p(x'_1, x'_2, ..., x'_m) = \sum_{i=1}^M \log \hat{f}_h(x'_i)
$$
where $(x'_1, x'_2, ..., x'_m)$ are the features across different measurements. We present an evaluation result later in Sec.~\ref{sec:new-evaluation} to show multi-measurement bootstrapping significantly improves the health detection performance.
% \textbf{Noise Mitigation.} The way that we brush our teeth is not fixed resulting in random noise added to the audio measurements. Therefore, we collect multiple audio samples for the same tooth over time, and estimate the joint distribution of these samples to average out the noise. We assume that these samples are independent, and as a result the joint distribution equals the multiplication of the marginal distributions estimated earlier using KDE.

% \textbf{Sample/Window Size.} Interestingly, there is a trade-off regarding the window size to construct a sample of features for a health tooth. As we increase the window size, we can form a sample of features that are robust to noise. On the other hand, as we decrease the window size, we can model each part of the same tooth separately, given that different parts may have different health conditions. Essentially, we can have a higher resolution modeling of a healthy tooth by tuning the window size parameter, and avoid averaging the features of the different areas of the same tooth, which may result in losing valuable health information.

\section{Mitigating error factors}
\label{sec:factors}

\kuang{In real-world scenarios, \name\ encounters various sources of error that can impede its performance. In this section, we first detail the noise suppression algorithm that was first mentioned in Sec.~\ref{subsec:vibration-model}. We then present the DTW-based sequence alignment approach for mitigating tooth matching errors.}

\subsection{Noise Suppression}
\label{subsec:noise-suppression}
\begin{figure}
\centering
% \vspace{-0.22in}
  \includegraphics[width=0.9\linewidth]{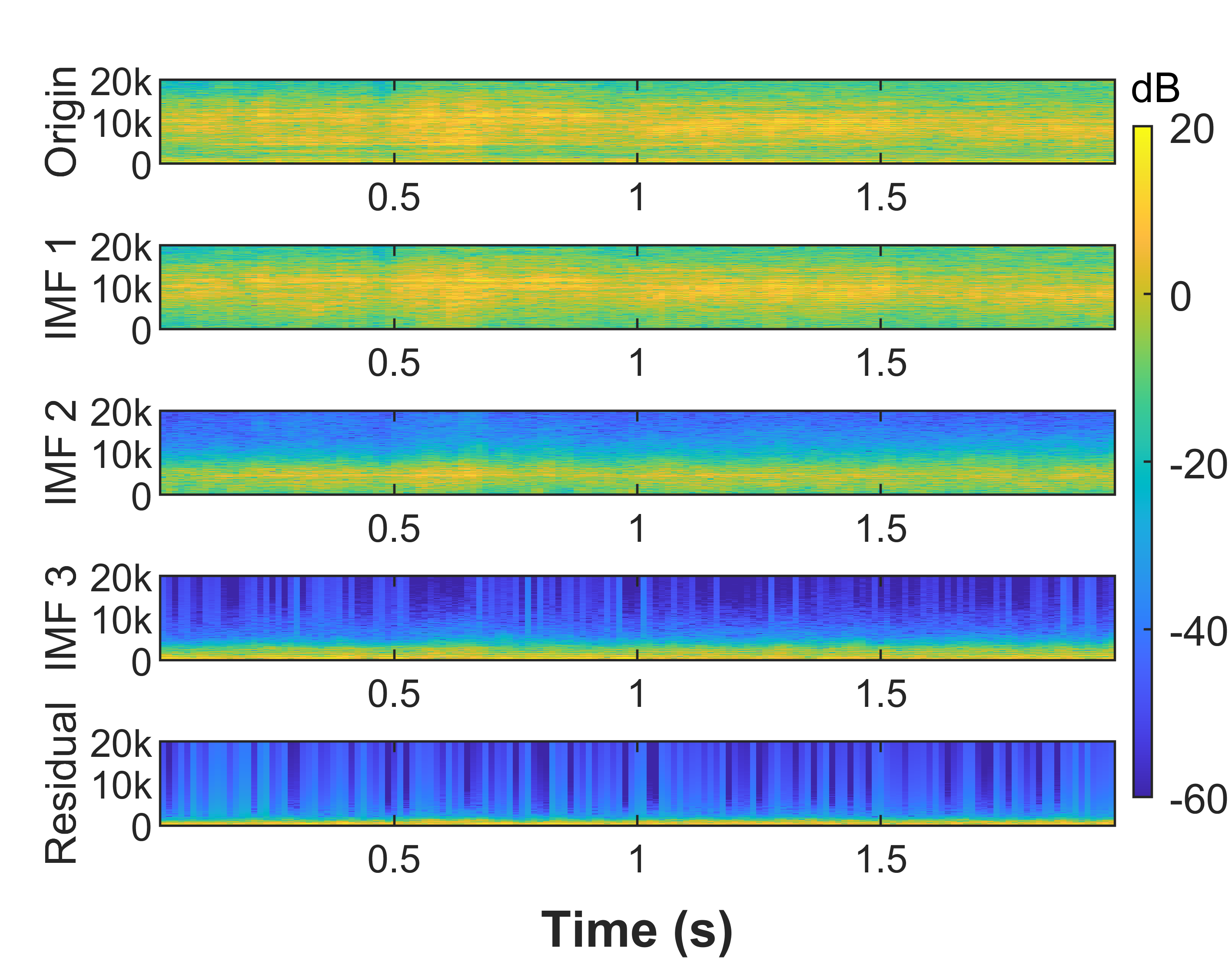}
    \vspace{-0.1in}
    \caption{Signal decomposition using EMD shown as spectrograms. IMF 1 and 2 correspond to resonance signals of teeth. IMF 3 corresponds to the excitation signal directly from the toothbrush, and the residual is environmental noise.}
    \vspace{-0.1in}
    \label{fig:emd}
\end{figure}

In real-world scenarios, the raw audio captured by the microphone is a mixture of various signals, including the resonance signal from the teeth, direct excitation signal from the toothbrush, environmental noise, and other artifacts. Although these components have different frequency distributions, they are likely to overlap in the frequency domain, making it harder to separate them simply through traditional band-pass filtering. As we modeled in Eq.~\ref{eq:whole-model}, the raw acoustic signal captured by the microphone $M(f)$ can be denoted as:
$$
    M(f) = Y(f) + D(f) + N(f)
$$
where $Y(f)$ is the response signal from the tooth, $D(f)$ is the direct path signal from the toothbrush, and $N(f)$ is other environmental noise. Only $Y(f)$ contains the tooth resonance behavior we desired which is required to be extracted for the later signature extraction step. To minimize the interference of other components and make the signal focus on $Y(f)$, we decompose the raw audio recording using Empirical Mode Decomposition~(EMD)~\cite{huang1998empirical}. EMD is a data-adaptive multi-resolution technique that decomposes signals into physically meaningful components. These components are called Intrinsic Mode Functions (IMFs) that have well-defined frequency ranges and amplitudes. Prior systems leverage EMD-based algorithms for heartbeats~\cite{zhang2020your, nfheart, sun2023earmonitor} or breathing monitoring~\cite{xu2019breathlistener} by decomposing and removing the noise terms such as motion artifacts. Fig.~\ref{fig:emd} demonstrates a result of decomposing an audio recording of teeth scanning into three IMFs and residual components. Based on our detailed empirical analysis, IMF-1 and IMF-2 are resonance signals from the teeth mainly in the frequency range of 4-20~kHz, and 2-7~kHz respectively, which is consistent with the observation that the tooth mainly resonants in 2-20~kHz (shown in Fig.~\ref{fig:sig-mic}). The IMF-3 is mainly the excitation signal directly from the toothbrush under 3~kHz ($D(f)$ in Eq.~\ref{eq:whole-model}), and the remaining residual component is mainly environmental noise (e.g.~surrounding speech) and other artifacts under 1~kHz. Thus, to perform noise suppression, \name\ only retains the first two IMF components and adds them up for further processing. We note that this suppression is done once on the entire original audio signal, rather than the individual 50-millisecond time chunks. 

As a result of eliminating residuals and IMF-3, the environmental noise in the raw audio signal is suppressed and the $D(f)$ term gets filtered out (detailed evaluation in Sec.~\ref{subsec:benchmark}). At the end of this process, we obtain the processed signal as:
\begin{equation}
   M_p(f) =  \sum_{i=1,2}IMF_{i}(M(f)) \approx Y(f) = H(f)B(f){X}(f)
\label{eq:emd-processed}
\end{equation}

\subsection{Sequence Alignment}\label{sec:alignment}

\begin{figure}
% \vspace{-0.03in}
  \includegraphics[width=\linewidth]{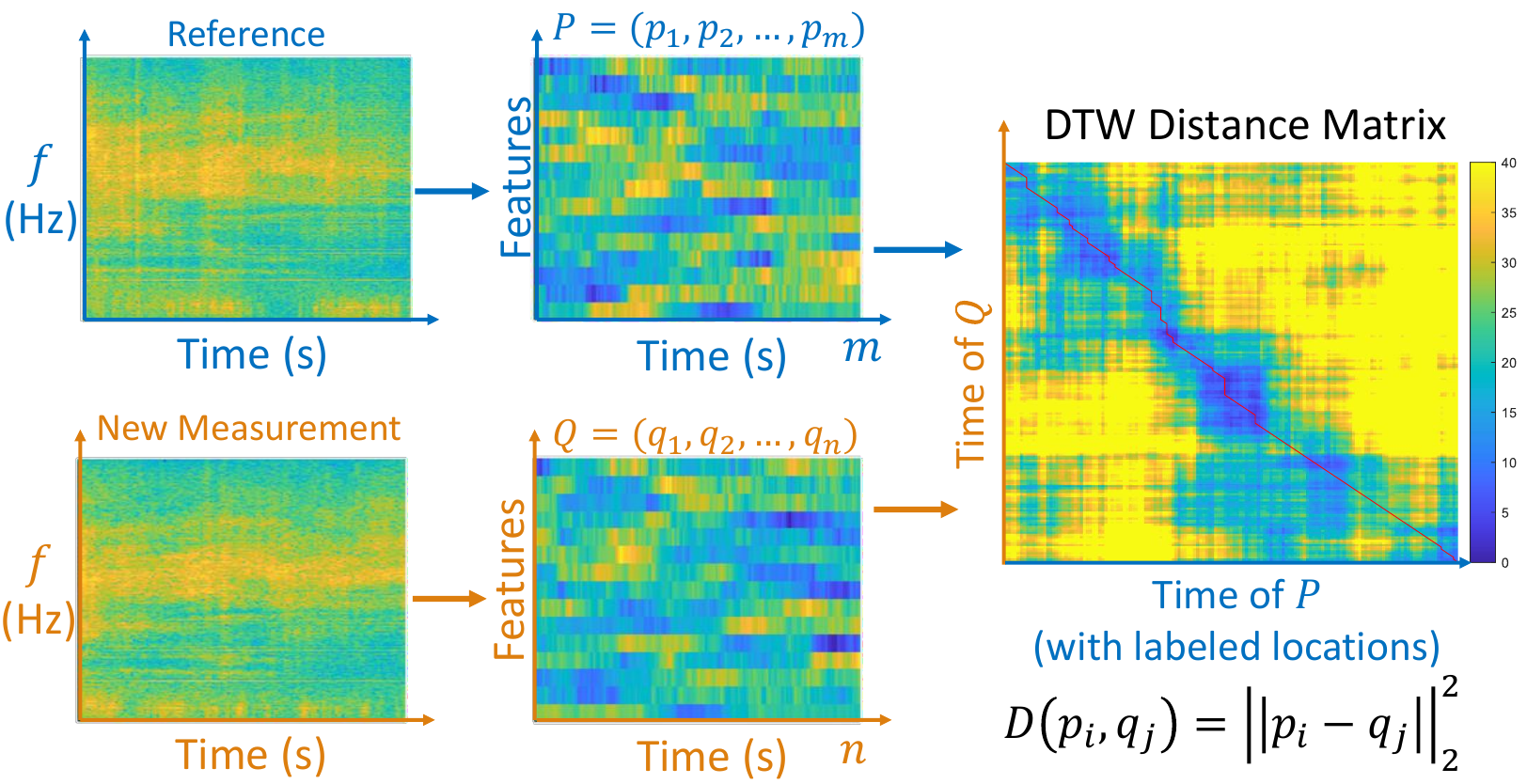}
    \vspace{-0.2in}
    \caption{\name\ performs sequence alignment using DTW based on the pair-wise distances between extracted features.}
    \vspace{-0.1in}
    \label{fig:aligner}
\end{figure}

\begin{figure*}
  \centering
  % \hspace{-0.1in}
  \begin{subfigure}[b]{0.13\linewidth}
    \includegraphics[width=\textwidth]{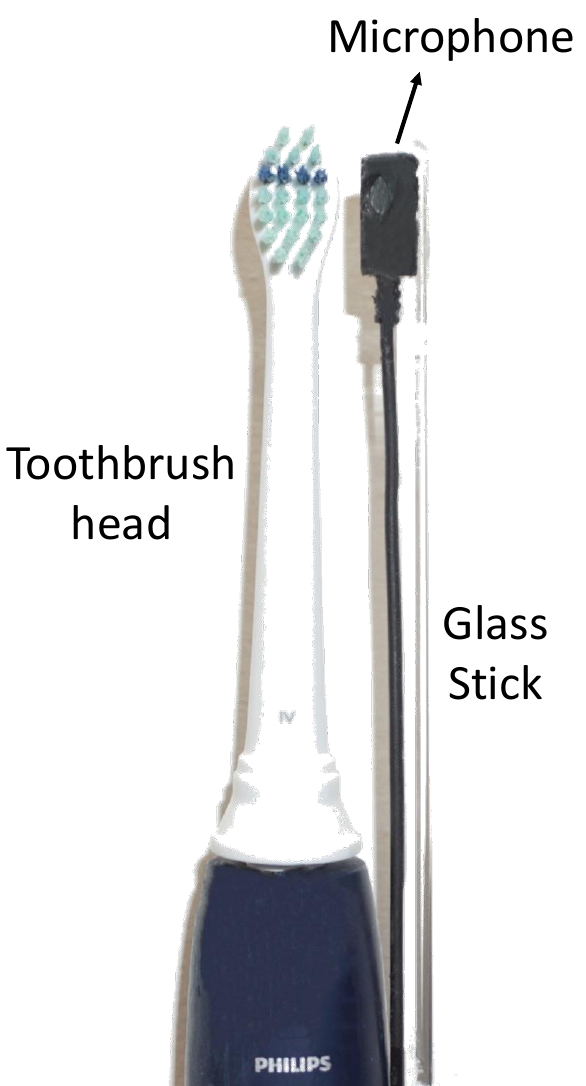}
    % \vspace{-0.2in}
    \caption{Hardware}
    \label{fig:hardware}
  \end{subfigure}
  \hspace{0.1in}
  \begin{subfigure}[b]{0.265\linewidth}
    \includegraphics[width=\textwidth]{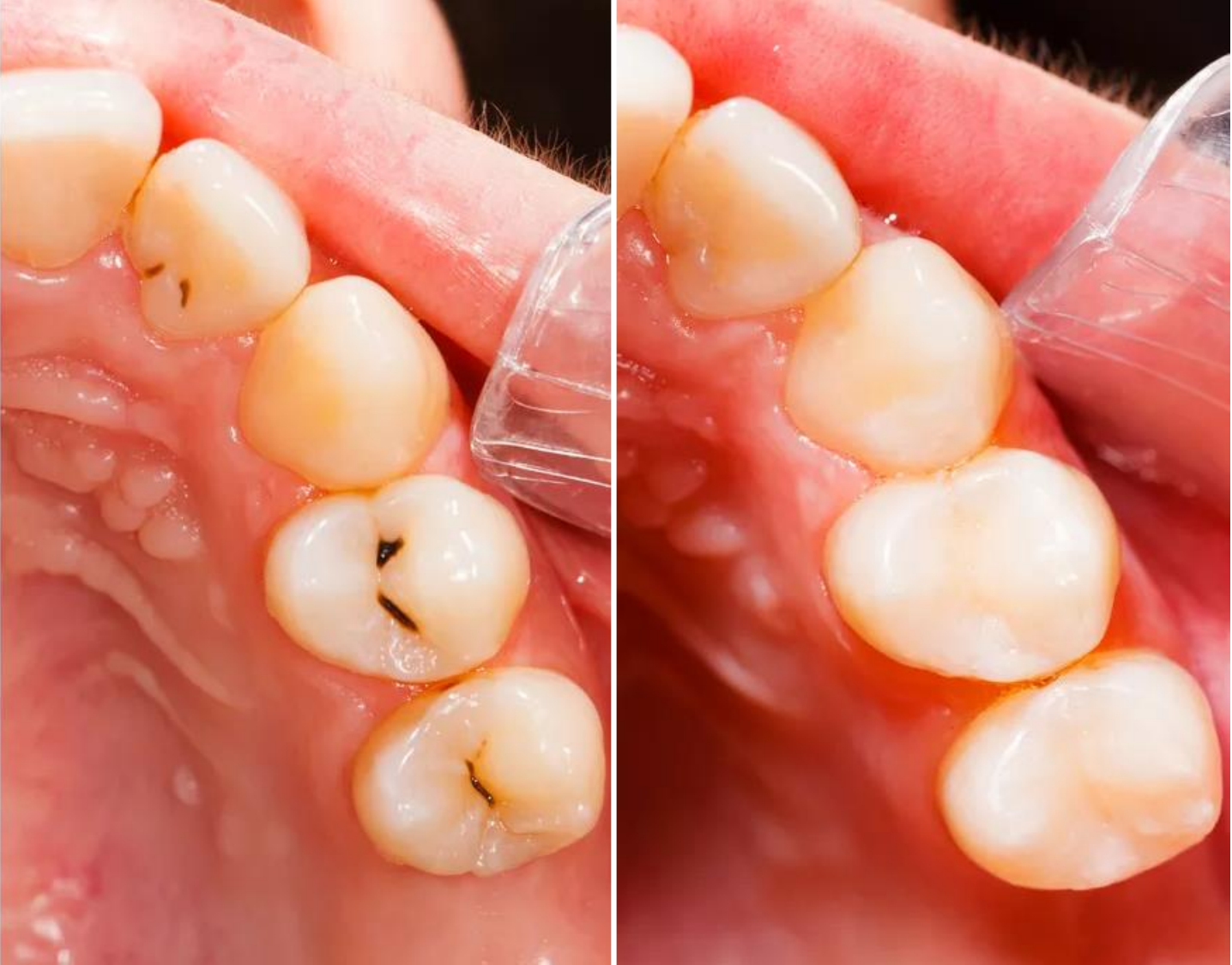}
    % \vspace{-0.2in}
    \caption{Dental Caries}
    \label{fig:caries-photo}
  \end{subfigure}
  \hspace{0.1in}
    \begin{subfigure}[b]{0.302\linewidth}
    \includegraphics[width=\textwidth]{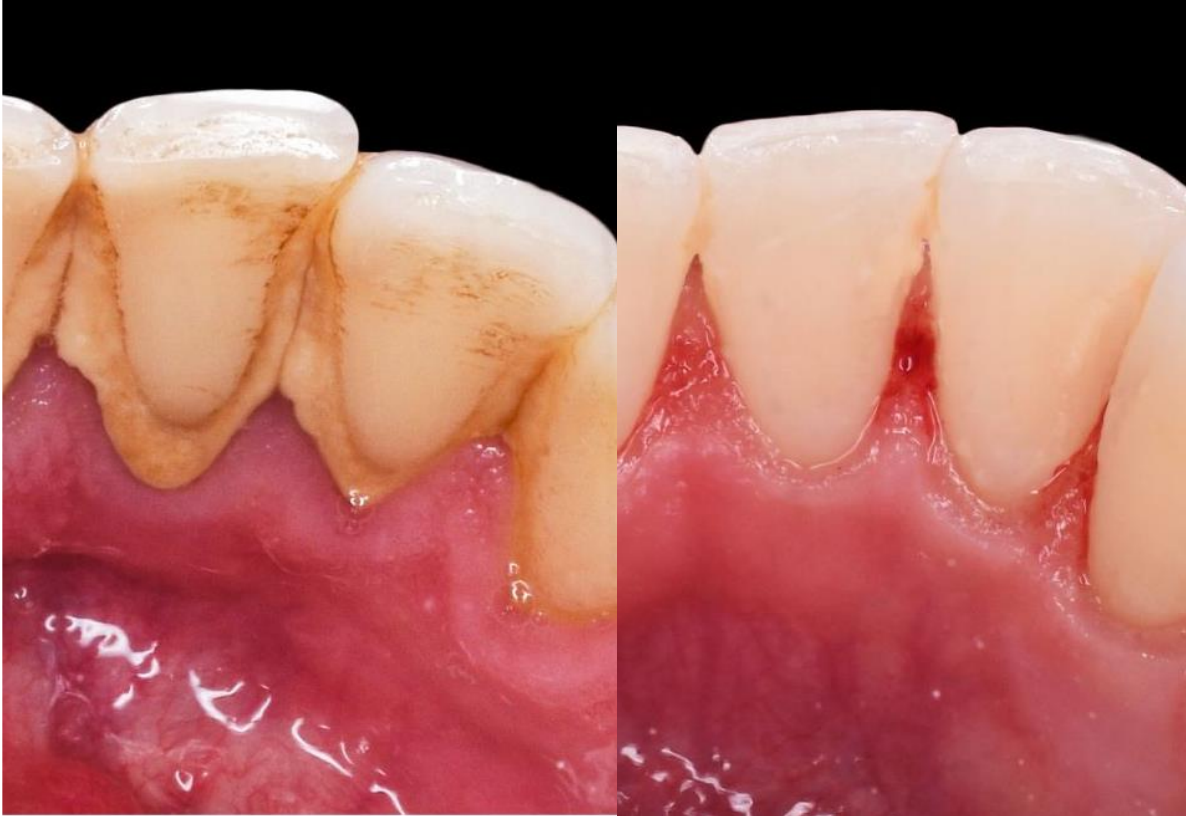}
    % \vspace{-0.2in}
    \caption{Dental Calculus}
    \label{fig:calculus-photo}
  \end{subfigure}
  \hspace{0.1in}
  \begin{subfigure}[b]{0.215\linewidth}
    \includegraphics[width=\textwidth]{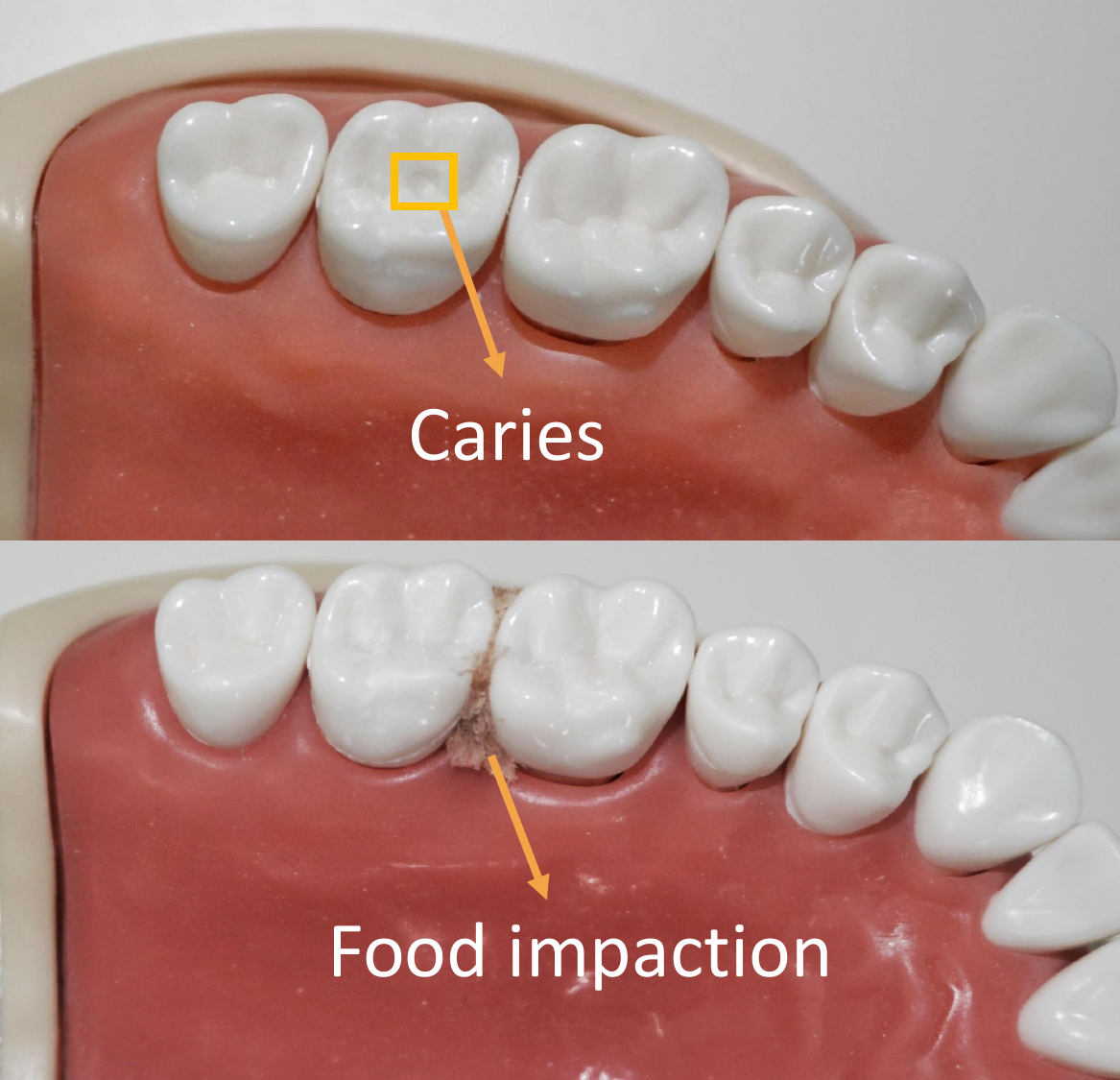}
    % \vspace{-0.2in}
    \caption{Model Emulation}
    \label{fig:model-photo}
  \end{subfigure}
  \vspace{-0.1in}
  \caption{(a) shows the hardware setup of \name. (b) and (c) are examples of dental caries and calculus before and after the treatment, which are two diseases we seek to detect in the user study in the dental clinic center. (b) shows a dental model we used to emulate caries and food impaction.}
  % \vspace{-0.1in}
  \label{fig:implementation}
\end{figure*}

\kuang{\name\ instructs the users to brush their teeth in a pre-specified order, guided by a video, to collect signatures from individual teeth. However, given the low-light conditions at some blind spots inside the mouth (such as inner upper molars), occasional matching errors can occur when a user collects measurements in these regions with poor visibility. For instance, a user may inadvertently miss brushing a tooth located in a blind spot. To limit the impact of matching errors in blind spots,  \name\ develops a sequence alignment algorithm based on Dynamic Time Warping (DTW) to align brushing sequences of the user's tooth with the references from a same quadrant of teeth, which allows to obtain signatures from each corresponding tooth more accurately.}

\kuang{In Sec.~\ref{sec:signature}, we obtain a tooth resonance signature in every 50~ms time window in the audio recording. Thus, as shown in Fig.~\ref{fig:aligner}, a time-series sequence of signatures is generated from the raw audio recording, which includes signatures from every individual tooth, as brushing is performed. The goal of the sequence alignment is to find the correct match of each signature to the corresponding tooth number (i.e. tooth location). To achieve this, \name\ aligns the received acoustic testing sequence produced from the scan with prior reference sequences obtained in dental clinics, where the locations of teeth brushed are known (e.g. via camera imaging). In this manner, it can isolate acoustic signatures of specific individual teeth. Note that \name\ only performs alignments for the sequences from the same quadrant of the same user.}

\kuang{\name\ seeks to accurately align the sequence even under the condition that the tooth signatures are changed by health conditions. Therefore, to make the sequence alignment resilient to health condition changes and other residual environmental noise, \name\ applies the same feature extraction algorithm as discussed in Sec.~\ref{sec:feature} to select a part of the signature as the input of the sequence alignment, instead of using the whole signature or raw audio recording. The features are selected by maximizing the separability between different teeth, and can best distinguish the location of the tooth brushed.}

Essentially, \name\ aims to align two time-series (the reference and the testing sequences) which may each involve brushing at a different speed (i.e. spending different durations on different teeth). We use Dynamic Time Warping (DTW), a standard technique known in time-series alignment (classically used in speech processing~\cite{myers1980performance, sakoe1978dynamic}), to match and find similarities between two temporal sequences.  For the distance metric in DTW, we use the square of Euclidean distance to evaluate the similarity between two feature samples. Specifically, for each pair of features in the two sequences $P = (p_1, p_2, ... p_m), Q = (q_1, q_2, ..., q_n)$, we obtain a distance:
$$D(p_i, q_j) = ||p_i-q_j||_2^2$$
Note that we perform normalization for each dimension of features using mean and standard deviation before the alignment. Fig.~\ref{fig:aligner} shows an example of the distance matrix for two sequences and the curve in red indicates the alignment result. As the ground truth of the tooth location corresponding to each point in the reference sequence is known, we can map the points in the testing sequence to the corresponding tooth locations based on the alignment curve.

% Note that during brushing, the toothbrush head may stay in a single tooth for up to a few seconds rather than only a ??single 50~ms time window.
After obtaining the tooth locations for each 50 ms time window, we can aggregate the adjacent signatures from the same tooth in the sequence by averaging these values to improve signal-to-noise ratio. Thus, the user can spend a longer period on the tooth they suspect is unhealthy to obtain a more accurate detection result. We further present an evaluation in Sec.~\ref{sec:new-evaluation} to show the relationship between the evaluation accuracy and the duration of time the brush remains on an individual tooth.

\section{Implementation and Evaluation}
\label{sec:new-evaluation}

\begin{figure*}
  \centering
  % \hspace{-0.1in}
  \begin{subfigure}[b]{0.25\linewidth}
    \includegraphics[width=\textwidth]{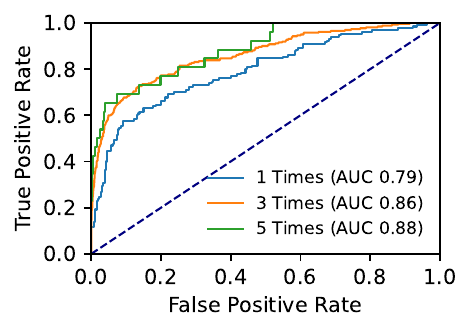}
    \vspace{-0.25in}
    \caption{\hspace{0mm}Food impaction (User Study)}
    \label{fig:food-user}
  \end{subfigure}
  \hspace{-0.1in}
  \begin{subfigure}[b]{0.25\linewidth}
    \includegraphics[width=\textwidth]{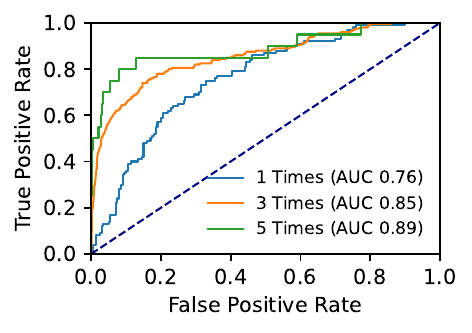}
    \vspace{-0.25in}
    \caption{\hspace{1mm}Food Impaction (Model)}
    \label{fig:food-model}
  \end{subfigure}
  \hspace{-0.1in}
  \begin{subfigure}[b]{0.25\linewidth}
    \includegraphics[width=\textwidth]{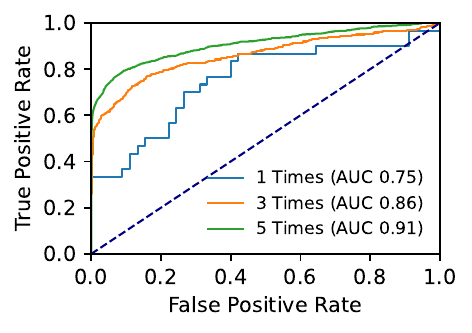}
    \vspace{-0.25in}
    \caption{\hspace{1mm}Caries (User Study)}
    \label{fig:caries-user}
  \end{subfigure}
  \hspace{-0.1in}
  % \newline
  \begin{subfigure}[b]{0.25\linewidth}
  % \vspace{-0.05in}
    \includegraphics[width=\textwidth]{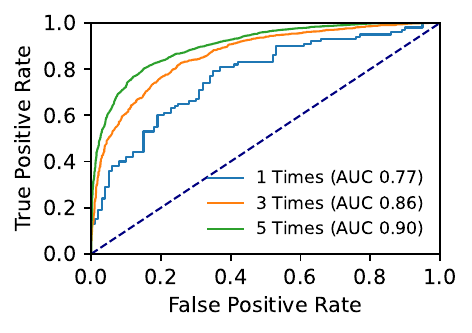}
    \vspace{-0.25in}
    \caption{\hspace{1mm}Caries (Model)}
    \label{fig:caries-model}
  \end{subfigure}
  \hspace{-0.1in}
  % \newline
  \begin{subfigure}[b]{0.26\linewidth}
  % \vspace{-0.1in}
    \includegraphics[width=\textwidth]{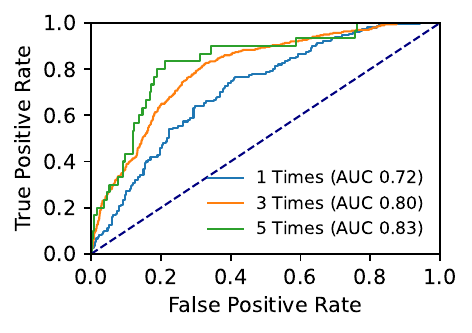}
    \vspace{-0.2in}
    \caption{\hspace{1mm}Calculus (User Study)}
    \label{fig:calculus-user}
  \end{subfigure}
  \hspace{-0.1in}
  \begin{subfigure}[b]{0.3\linewidth}
      \centering
      % \vspace{-0.25in}
      \includegraphics[width=\textwidth]{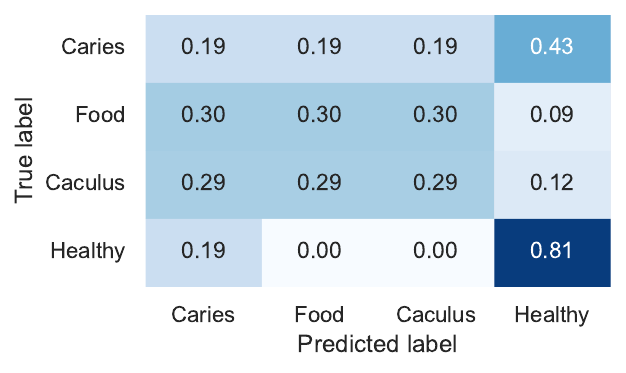}
      \vspace{-0.2in}
      \caption{\hspace{2mm}Without Feature Selection}
      \label{fig:confusion-matrix-wo}
  \end{subfigure}
% \hspace{-0.1in}
  \begin{subfigure}[b]{0.3\linewidth}
      \centering
      % \vspace{-0.25in}
      \includegraphics[width=\textwidth]{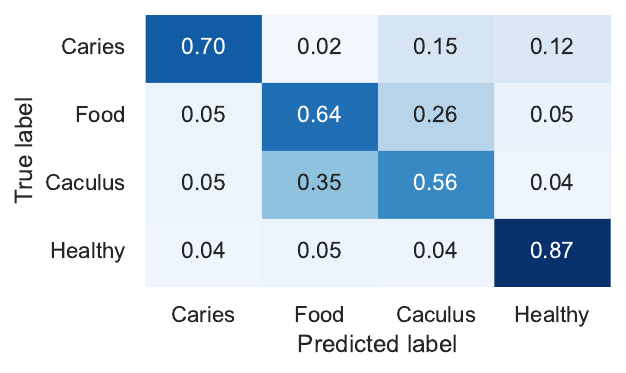}
      \vspace{-0.2in}
      \caption{\hspace{2mm}With Feature Selection}
      \label{fig:confusion-matrix-w}
  \end{subfigure}
  \vspace{-0.1in}
  \caption{\name\ Results: (a) - (e) show the  ROC curves of the three health detection tasks. Three solid lines colored in blue, orange, and green show the detection performance using different number of cumulative measurements. The blue dashed lines show the performance of random choosing.  \feb{(f) and (g) shows the confusion matrix of discriminating different diseases without and with the feature selection algorithm.}}
  % \vspace{-0.1in}
  \label{fig:overall-results}
\end{figure*}

We implement the prototype of \name\ on a Philips Sonicare ProtectiveClean 6100 electric toothbrush~\cite{sonicare6100} and a Voice Technologies VT500X waterproof microphone~\cite{vt500x}. For toothbrush heads, we choose Brushmo Compact Replacement Heads~\cite{brushmo} as it has a smaller size compared with the standard one. As shown in Fig.~\ref{fig:hardware}, the microphone is fixed 1.25~cm close to the toothbrush head using a tiny glass stick, which is integrated onto the toothbrush body using a hot-melt adhesive. The microphone is sampling at 44.1~kHz for audio recording and connected to a laptop through a 3.5~mm audio jack. For signal processing, we choose a 50 milliseconds time window and 75\% overlap to perform STFT. For signals from users and dental models, we empirically choose the frequency range of 2-16~kHz, and 2-18~kHz respectively to obtain the cepstrum (illustrated in Sec.~\ref{subsec:extract-alg}). \feb{We note that the only learnable parameter in our pipeline is the selected feature index presented in Sec.~\ref{sec:feature}. To evaluate the generalizability of our method, the following evaluation result is presented by performing the leave-one-out cross-validation in the feature selection module.}
% A video camera is put in front of user's mouth to capture the real-time ground truth locations of the toothbrush. Note that the camera is only needed as ground truth for the evaluation of sequence alignment -- \name\ does not require or rely on camera data. We use an online video labeling tool~\cite{Labelbox} to label the tooth being brushed in each video frame.

In the following sections, we first present the technical evaluation of \name, followed by an expert evaluation by interviewing two experienced dentists.

% In the following sections, we present the evaluation results of \name\ in three categories: (1) \textbf{Health Detection}: Reports \name's accuracy in detecting the three identified dental conditions. (2) \textbf{Sequence Alignment}: Presents the accuracy to identify the correct tooth number when using time-series alignment to mitigate the matching error. (3) \textbf{Expert Evaluation:} An interview with two expert dentists.

\subsection{Technical Evaluation}
In the evaluation in this section, we focus on the capabilities of \name\ to detect three dental conditions: dental caries, dental calculus, and interdental food impaction. \kuang{For the main result section (Sec. \ref{sec:mainres})}, we ensure that the tooth being brushed is always the tooth of interest (i.e. no matching errors). We further evaluate the capability of \name\ to mitigate the matching errors in Sec.~\ref{subsec:sequence-alignment}.

\subsubsection{Data Collection Procedure}\leavevmode

\sssec{Evaluation through user study:} We conducted an IRB-approved user study on 19 participants (12 males and 7 females). We evaluated the health detection for caries and calculus in a dental clinical center, where we perform measurements on real patients under the guidance of professional dentists. Each measurement is taken by brushing from the chewing surface of the tooth and lasts for 3-4 seconds. For the evaluation of caries detection, we took measurements on specific teeth with caries (labeled by dentists) before and after the dental fillings (example shown in Fig.~\ref{fig:caries-photo}). For the evaluation of calculus detection, we perform measurements on teeth originally with calculus before and after the dental cleaning (example in Fig.~\ref{fig:calculus-photo}). For every tooth of interest from every patient, we collected 10 measurements each before and after the treatments (20 measurements in total). We assume the teeth after the treatments are healthy, use a subset of samples (5 measurements) after the treatments as references, and test on the rest of the samples (15 measurements). We use random combinations of 3 or 5 measurements among the 15 measurements to evaluate the multi-measurement bootstrapping as presented in Sec.~\ref{sec:detection}.

We also evaluated detection for interdental food impaction in a home setting. We collected measurements on the original normal teeth, teeth with food impaction after eating, and the teeth after flossing to clear out impacted food. We collected 10 measurements in each stage (30 measurements in total) for every tooth of interest from every participant. We use 5 measurements from the original normal teeth as references and test on the others.

\sssec{Evaluation on dental models:} \rr{In order to perform a comprehensive evaluation of our system in more controlled and flexible settings, such as evaluating under different brushing strengths, brushing duration, and damage to different parts of teeth, we choose to increase the scale of the evaluation by using dental models, mainly due to ethical reasons (i.e. avoiding possibility of damage to human teeth). We note that the properties of teeth material on dental models are not equivalent to those on humans. By collaborating with expert dentists, we select the dental models with as accurate as possible structures that are designed for professional dental training (e.g. filling, cleaning).} \feb{Our following evaluation results demonstrate consistent health detection performance on user study and dental model, which validate the effectiveness of involving dental models into our experiments. } 
% We further increase the scale of the evaluation by using professional dental models designed for dental training. We perform this evaluation to allow for more controlled damage to be performed on the teeth, unlike a relatively more unstructured user study context. 

To emulate the diseases, we use an electric drill to produce small holes on the surface of the model's teeth to emulate caries, and insert meat samples in between two teeth to emulate food impaction (shown in Fig.~\ref{fig:model-photo}). We collect measurements on the model teeth before and after the changes for our evaluation. Each measurement is taken by brushing from the chewing surface of the tooth and lasts for 5-6 seconds. The number of measurements on each tooth of interest is the same as in the user study.

\begin{figure*}[t]
% \begin{minipage}{0.48\textwidth}
    \centering
    % \vspace{0.11in}
    \begin{subfigure}[b]{0.25\linewidth}
% \vspace{-0.1in}
    \includegraphics[width=\textwidth]{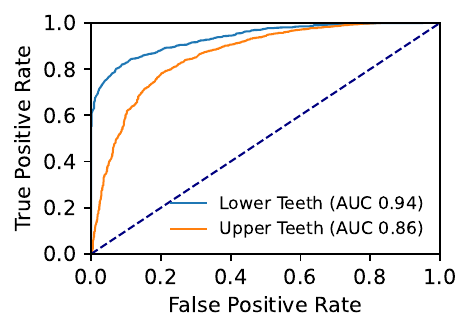}
    \vspace{-0.2in}
    \caption{\hspace{1mm}Lower vs. Upper Teeth}
    \label{fig:lowerVsUpper}
  \end{subfigure}
  \hspace{-0.1in}
     \begin{subfigure}[b]{0.25\linewidth}
  % \vspace*{-1.2in}
    \includegraphics[width=\textwidth]{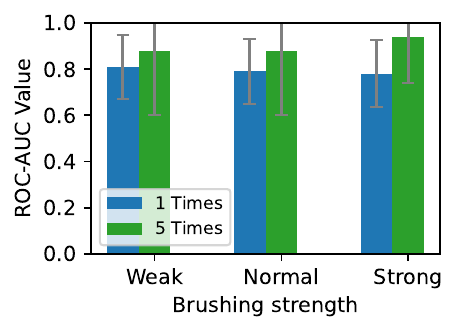}
    \vspace{-0.25in}
    \caption{\hspace{1mm}Brushing strength}
    \label{fig:strenght}
  \end{subfigure}
  \hspace{-0.1in}
  \begin{subfigure}[b]{0.25\linewidth}
    \includegraphics[width=\textwidth]{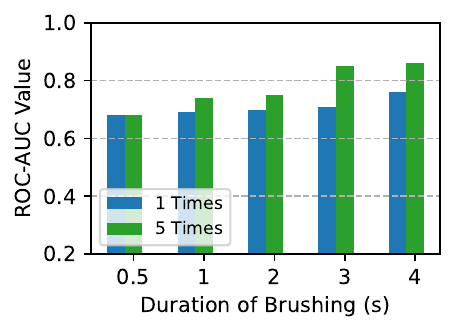}
    \vspace{-0.25in}
    \caption{\hspace{1mm}Duration of brushing}
    \label{fig:testlength}
  \end{subfigure}
  \hspace{-0.1in}
  \begin{subfigure}[b]{0.25\linewidth}
    \includegraphics[width=\textwidth]{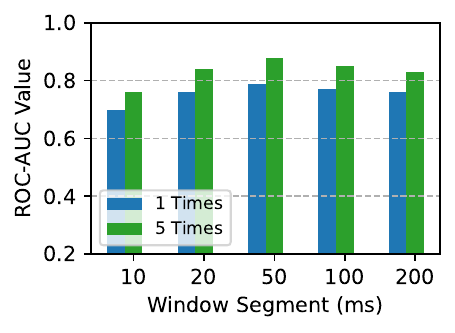}
    \vspace{-0.25in}
    \caption{\hspace{1mm}Window Size}
    \label{fig:winlength}
  \end{subfigure}
\vspace{-0.1in}
  \caption{Benchmarking results: (a) shows the detection accuracy on lower teeth v.s. upper teeth. (b) compares the detection accuracy under different brushing strengths. (c) compares the detection accuracy of different durations of brushing on a single tooth. \feb{(d) compares the detection accuracy when using different window sizes.}}
  % \vspace{-0.1in}
% \end{minipage}
% \hspace{0.1in}
% \begin{minipage}{0.48\textwidth}

% \end{minipage}
% \vspace{-0.2in}
\end{figure*}

\subsubsection{Main Results of Health Detection}\label{sec:mainres}\leavevmode

We evaluate the health detection performance of \name\ using Receiver Operating Characteristic~(ROC) curve, which shows the detection sensitivity (True Positive Rate) versus False Positive Rate by varying the decision threshold. The Area Under Curve (AUC) value of the ROC curve represents \name's overall capability to distinguish unhealthy samples from healthy samples. 
% We note that each measurement taken on the tooth of interest only lasts for 3-4~s, and 5-6~s for user study and dental model experiments respectively.

\sssec{Interdental Food Impaction:} We evaluate \name\ for the detection of interdental food impaction through the user study and on the dental model. We collected measurements on 26 teeth in total in the user study. As shown in Fig.~\ref{fig:food-user}, using a single measurement for around 3-4 seconds, \feb{\name\ achieves a 0.79 AUC value with a 95\% confidence interval of (0.743, 0.844). With 3 and 5 measurements, \name\ bootstraps the performance to 0.86~(0.833, 0.891), and 0.848~(0.799, 0.968) respectively.}
% \name\ achieves a 0.76 AUC value with a 95\% confidence interval of (0.702, 0.809). With 3 and 5 measurements, \name\ bootstraps the performance to 0.81~(0.783, 0.847), and 0.84~(0.744, 0.935) respectively.

We collected measurements on 20 dental model teeth with food impaction. As shown in Fig.~\ref{fig:food-model},  \name\ achieves a 0.76~(0.699, 0.821) AUC value using a single measurement. With three and five measurements, \name\ bootstraps the performance to 0.85~(0.821, 0.888), and 0.89~(0.792, 0.982) respectively. \feb{We observe that the detection performance we obtain from user study and dental model emulation is very close to each other.}
% We note that the user study accuracy is slightly lower than the experiments on the model under multiple measurements mainly because some food is partially cleaned out during the data collection procedure.

\sssec{Caries Detection:} We evaluate \name\ for caries detection through the user study and on the dental model. We collected data on five teeth with caries from real patients during the user study and the results are shown in Fig.~\ref{fig:caries-user}. \name\ achieves 0.75~(0.637, 0.870), 0.86~(0.834,0.884), and 0.91~(0.893, 0.922) AUC values with 1, 3, and 5 accumulative measurements respectively. Similarly, we collected data on 10 teeth with caries on the dental model the results are shown in Fig.~\ref{fig:caries-model}, where it achieves a performance very close to the user study -- 0.77~(0.704, 0.835), 0.86~(0.848, 0.878), and 0.90~(0.894, 0.912) AUC values under 1, 3, and 5 measurements respectively.

\sssec{Caculus Detection:} We evaluate \name\ for caries detection only through the user study. We collected data on 30 teeth with calculus from real patients and the results are shown in Fig.~\ref{fig:calculus-user}. \name\ achieves 0.72~(0.663,0.768), 0.80~(0.768, 0.829), and 0.83~(0.738,0.920) AUC values with 1, 3, and 5 accumulative measurements respectively. Note that the dental calculus on the users who get dental cleaning frequently is generally very tiny, which induces smaller changes to the tooth resonances compared with caries or food impaction, so the detection performance is slightly lower.

\sssec{Accuracy vs. related work:}  OralCam~\cite{OralCam} is the only related work to \name\ that targets low-cost in-home dental health monitoring. OralCam reports 0.84, 0.79 AUC values for caries and calculus detection, which is close to the performance of \name\ using 3 measurements (0.86, 0.80 AUC), but lower than our results with 5 measurements (0.91, 0.83 AUC). We note OralCam evaluates on a visible disease dataset, but \name\ also has the unique advantage of detecting diseases at the most inner molars or even invisible corners where common cameras can hardly capture high-quality images.

\sssec{Capability to Discriminate Different Diseases:} \feb{Besides detecting each individual disease, \name\ is also able to discriminate between different diseases attributed to the feature selection algorithm presented in Sec.~\ref{sec:feature}. We evaluate the performance of \name\ to discriminate different diseases by combing the collected datasets from the user study. We first present the result without using the feature selection algorithm in a confusion matrix shown in Fig.~\ref{fig:confusion-matrix-wo}. We input the extracted tooth signature as a whole (green part in Fig.~\ref{fig:cep-1-cep}) into the health detection module, and consider all three diseases to have equal probabilities for the positive outputs from the health detection. While the true negative rate is relatively high (0.81), the system can not distinguish different diseases without specialized features.}

\feb{The confusion matrix shown in Fig.~\ref{fig:confusion-matrix-w} demonstrates the performance of \name\ to distinguish different diseases when the feature selection algorithm is integrated. We can observe that food impaction and calculus are more challenging for the system to discriminate, since both of them predominantly occur in interdental spaces, inducing changes of the resonance behaviors of the teeth in a similar way. We note that the results are generated by picking operating points on the ROC curves to maximize the overall accuracy. In actual deployment, \name\ can choose a more conservative operating point to further reduce the false positive rate.}

% \sssec{Capability to Discriminate Different Diseases:} \rr{We also evaluate the performance of \name\ to discriminate different diseases by combing the collected datasets from the user study. We pick operating points on each of the ROC curves of three different diseases separately~(Fig.~\ref{fig:food-user}, \ref{fig:caries-user}, \ref{fig:calculus-user}) to maximize the classification accuracy in total. The result shown in Fig.~\ref{fig:confusion-matrix} demonstrates the performance of \name\ to distinguish different diseases using only single measurement. Such a capability is mainly attributed to the feature selection algorithm presented in Sec.~\ref{sec:feature}. We can observe that food impaction and calculus are more challenging for the system to discriminate, since both of them predominantly occur in interdental spaces, inducing changes of the resonance behaviors of the teeth in a similar way. While the results presented in Fig.~\ref{fig:confusion-matrix} are generated under the system objective of maximizing the overall accuracy, in actual deployment, \name\ can choose a more conservative operating point to further reduce the false positive rate. }

\begin{figure*}[t]
  \centering
  % \hspace{-0.1in}
  \begin{subfigure}[t]{0.23\linewidth}
    \includegraphics[width=\textwidth]{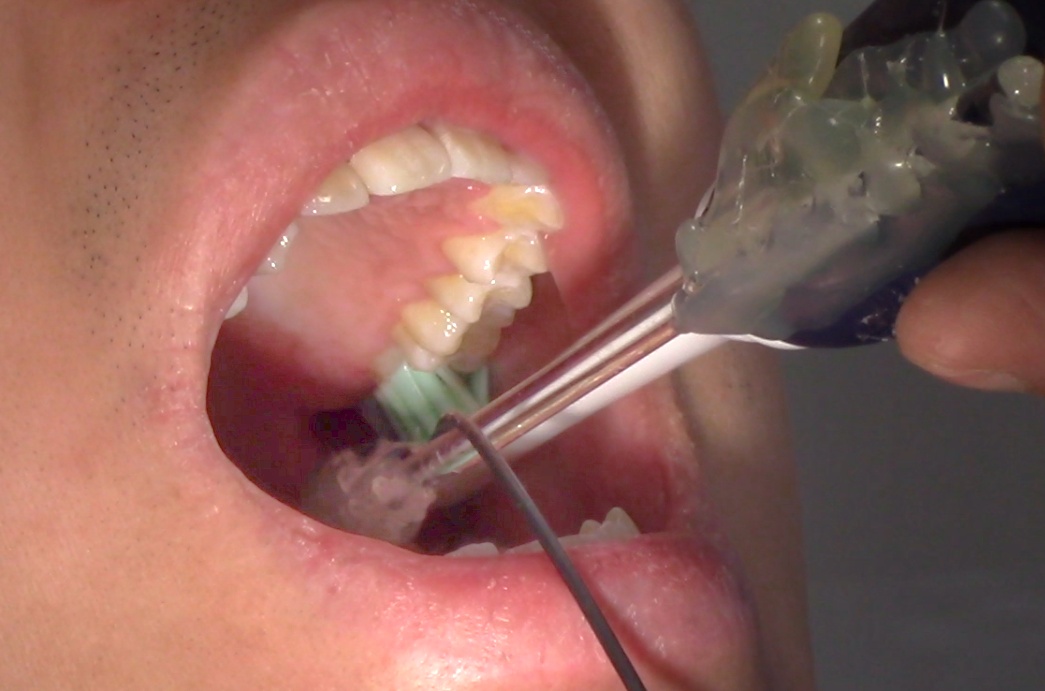}
    % \vspace{-0.2in}
    \caption{\hspace{1mm}A video frame}
    \label{fig:data-labeling}
  \end{subfigure}
  \hspace{0.05in}
  \begin{subfigure}[t]{0.25\linewidth}
    \includegraphics[width=\textwidth]{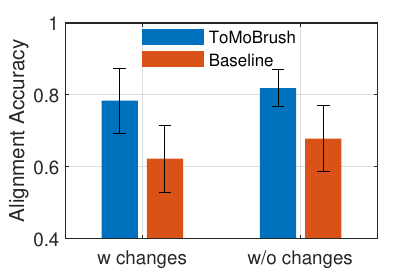}
    % \vspace{-0.2in}
    \caption{\hspace{1mm}Alignment accuracy}
    \label{fig:location-accuracy}
  \end{subfigure}
  \hspace{-0.1in}
  \begin{subfigure}[t]{0.25\linewidth}
    \includegraphics[width=\textwidth]{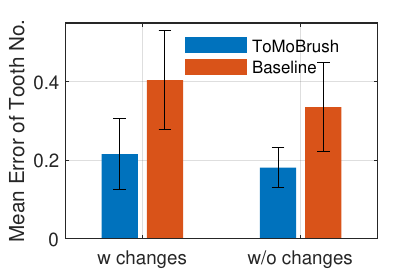}
    % \vspace{-0.2in}
    \caption{\hspace{1mm}Alignment Error}
    \label{fig:location-error}
  \end{subfigure}
  \hspace{-0.1in}
  \begin{subfigure}[t]{0.26\linewidth}
    % \vspace*{-1.2in}
    \includegraphics[width=\textwidth]{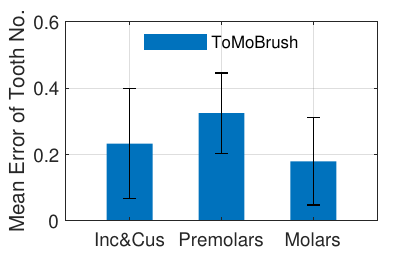}
    % \vspace{-0.1in}
    \caption{\hspace{0mm}Types of Teeth}
    \label{fig:loc-parts}
  \end{subfigure}
  \vspace{-0.1in}
  \caption{Evaluation results of sequence alignment. (a) shows an example of a recorded video frame to capture the ground-truth location of the toothbrush. (b) shows the accuracy of the sequence alignment to the correct tooth number. (c) shows the average alignment error of tooth number. (d) shows the alignment error on different types of teeth (Inc\&Cus: Incisors and Cuspids). }
  % \vspace{-0.1in}
  \label{fig:overall-results}
\end{figure*}

% \begin{figure*}
%   \centering
%   % \hspace{-0.1in}
%   \begin{subfigure}[b]{0.25\linewidth}
%     \includegraphics[width=\textwidth]{figures/strength.pdf}
%     \vspace{-0.2in}
%     \caption{\hspace{1mm}Brushing strength}
%     \label{fig:strenght}
%   \end{subfigure}
%   \hspace{-0.1in}
%   \begin{subfigure}[b]{0.25\linewidth}
%     \includegraphics[width=\textwidth]{figures/testlength_new.pdf}
%     \vspace{-0.2in}
%     \caption{\hspace{1mm}Duration of brushing}
%     \label{fig:testlength}
%   \end{subfigure}
%   \hspace{-0.1in}
%       \begin{subfigure}[b]{0.25\linewidth}
%     \includegraphics[width=\textwidth]{figures/emd-1time.pdf}
%     \vspace{-0.2in}
%     \caption{\hspace{1mm}1-Time Measurement}
%     \label{fig:emd-1time}
%   \end{subfigure}
%   % \hspace{-0.15in}
%   \begin{subfigure}[b]{0.25\linewidth}
%     \includegraphics[width=\textwidth]{figures/emd-5time.pdf}
%     \vspace{-0.2in}
%     \caption{\hspace{1mm}5-Times Measurements}
%     \label{fig:emd-5time}
%   \end{subfigure}
%   % \vspace{-0.1in}
%   \caption{Evaluation results of different factors. (a) compares the detection accuracy under different brushing strengths. (b) compares the detection accuracy of different durations of brushing on a single tooth. (c)-(d) shows the detection accuracy with and without the denoising algorithms in Sec.~\ref{subsec:noise-suppression} under 1 and 5 times measurements. }
%   % \vspace{-0.1in}
%   \label{fig:overall-results}
% \end{figure*}
\subsubsection{Benchmarking on Experiment Setup}\leavevmode

\sssec{Upper vs. Lower Teeth:} We compare the detection accuracy of caries on the dental model for the upper teeth versus the lower teeth when using 5 times cumulative measurements. Fig.~\ref{fig:lowerVsUpper} shows \name\ achieves a better detection performance on lower teeth (AUC = 0.94) compared with upper teeth (AUC = 0.86). Brushing upper teeth is usually more challenging as the applied force is in the opposite direction of gravity. As a result, the brush head generally has better contact with the lower teeth leading to better detection performance. 
% Fig.~\ref{fig:lowerVsUpper} shows significant improvement for the detection accuracy when moving from brushing the upper teeth (AUC = 0.86 AUC) to brushing the bottom teeth (AUC = 0.94). Brushing the upper teeth is usually challenging compared to the lower teeth as it's easier to see what are brushing for the lower teeth. Moreover, food debris fall off from the upper to the lower part of the mouth, resulting in clear changes for \name\ to detect.

\label{subsec:benchmark}
\sssec{Brushing Strength:} In this experiment, we study how the brushing strength affects the detection accuracy of food impaction detection on the dental model in Fig.~\ref{fig:strenght}. This study shows slight changes in accuracy when brushing strength varies, showing that \name\ is robust to different brushing behaviors. \kuang{As we discussed in Sec.~\ref{subsec:extract-alg}, our signature extraction algorithm models and isolates the effects of different brushing behaviors, such as strength and tiny movement. During our experiments, the three different levels of brushing strength are observable in the lower components of the cepstrum (an example shown in Fig.~\ref{fig:cep-1-low}). In our collected dataset, the total energy of the lower cepstral components when using normal strength is approximately 50\% of the one using strong strength, and the energy when using weak strength is approximately 10\% of the one using strong strength.}

Interestingly, the detection accuracy increases as the brushing strength increases when \name\ has enough measurements (5 times). As the strength increases, the brush bristles have closer and tighter contact with the brushed tooth, which in turn produces a response resonances with a higher signal-to-noise ratio.

\sssec{Brushing Duration:} We evaluate in this experiment how the duration of each brushing measurement on a single tooth affects the detection accuracy of food impaction in the user study. \name\ seeks to average the tooth signatures from the same tooth across multiple time windows to improve the signal-to-noise ratio. As Fig.~\ref{fig:testlength} shows the detection accuracy increases as the brushing duration increases for each tooth as expected, achieving relatively good performance when the duration reaches 3-4 seconds, which is also the chosen duration in our user study.

\sssec{Length of Window Segment:} \feb{We evaluate in this experiment how the configuration of window size affects the detection accuracy of food impaction in the user study. \name\ averages the signatures from the same tooth across time windows to improve the SNR. If the window size is too small, the frequency resolution of each window would be low which hinders the signature extraction algorithm from obtaining fine-grained signatures with detailed information. On the other hand, if the window size is too large, the number of windows staying on a single tooth would be small, thus the averaging can not improve the SNR significantly. The result in Fig.~\ref{fig:winlength} shows the detection accuracy when using different window sizes. The optimal window length in the experiment is 50 ms, which is also the chosen configuration in our main evaluation.}

\sssec{Different Sonic Toothbrushes:} \feb{We further replicate our prototype on two other sonic toothbrushes, Philips Sonicare 4100~\cite{sonicare4100} and Oralvue Seebrush~\cite{seebrush}. We evaluate the two prototypes for food impaction detection using dental models. Sonicare 4100 achieves 0.78 and 0.90 ROC-AUC values using 1 and 5 measurements respectively. Seebrush achieves 0.72 and 0.85 ROC-AUC values using 1 and 5 measurements respectively. The performance of both toothbrushes is within the margin of error of our main experiment prototype, which demonstrates the generalizability of our system to enable dental health sensing on different types of sonic toothbrushes.}

% \sssec{Types of Teeth:} In this experiment, we show how the different types of teeth, like Incisors, Cuspids, Premolars, and Molars, affect the alignment error. Fig.~\ref{fig:loc-parts} shows the mean error of detecting the tooth number that the user is currently brushing, and how it varies over different types of teeth. 
% Since the premolars are generally smaller and have irregular non-flat shapes compared with other teeth, it introduces slightly higher alignment errors. 
% As the molars are hard to reach and brush properly, it is also challenging to align them. This also similar to the incisors and cuspids, as it's challenging to brush each of them separately given their size is usually smaller than the brush head itself.

\subsubsection{Benchmarking on Error Mitigation Modules}\leavevmode
\label{subsec:sequence-alignment}

In this section, we present the evaluation results of the Noise Suppression module (Sec.~\ref{subsec:noise-suppression}) and the Sequence Alignment module (Sec.~\ref{sec:alignment}) to mitigate the errors.
\begin{figure}[h]
    \centering
% \vspace{-0.22in}
  \centering
  % \hspace{-0.1in}
  \begin{subfigure}[b]{0.5\linewidth}
  % \vspace{-0.1in}
    \includegraphics[width=\textwidth]{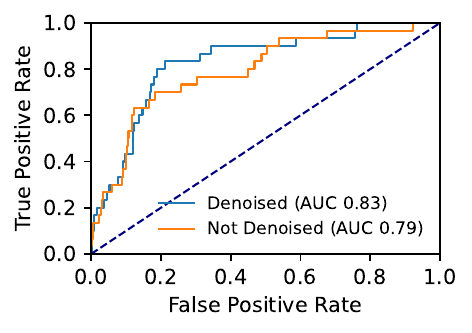}
    \vspace{-0.2in}
    \caption{Calculus (Quiet Room)}
    \label{fig:emd-calculus}
  \end{subfigure}
  \hspace{-0.1in}
  \begin{subfigure}[b]{0.5\linewidth}
  \vspace{-0.1in}
    \includegraphics[width=\textwidth]{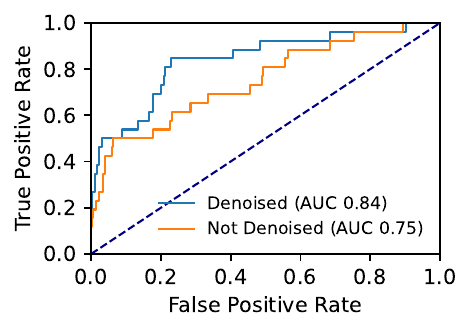}
    \vspace{-0.2in}
    \caption{Food Impaction (Noisy Room)}
    \label{fig:emd-food}
  \end{subfigure}
  \vspace{-0.1in}
  \caption{Noise Suppression Benchmark: The health detection performance in the user study with and without the denoised algorithms}
 \vspace{-0.1in}
  \label{fig:emd-eval}
\end{figure}

% \begin{figure*}[h]
%   \centering
%   % \hspace{-0.1in}
%   \begin{subfigure}[b]{0.3\linewidth}
%     \includegraphics[width=\textwidth]{figures/emd-1time.pdf}
%     \vspace{-0.2in}
%     \caption{\hspace{1mm}1-Time Measurement}
%     \label{fig:emd-1time}
%   \end{subfigure}
%   % \hspace{-0.15in}
%   \begin{subfigure}[b]{0.3\linewidth}
%     \includegraphics[width=\textwidth]{figures/emd-5time.pdf}
%     \vspace{-0.2in}
%     \caption{\hspace{1mm}5-Times Measurements}
%     \label{fig:emd-5time}
%   \end{subfigure}
%   \vspace{-0.1in}
%   \caption{Noise Suppression Evaluation: The health detection performance in the user study with and without the denoised algorithms}
%  % \vspace{-0.2in}
%   \label{fig:emd-eval}
% \end{figure*}

\sssec{Noise Suppression:} \feb{We illustrate in this experiment how the Noise Suppression module (Sec.~\ref{subsec:noise-suppression}) affects \name. We present the accuracy of the calculus and food impaction detection in our user study. The user study of dental calculus detection is conducted in a relatively quiet dental clinical room, while the user study of food impaction is conducted in a relatively noisy home setting with different noise sources, such as surrounding speech, air conditioner humming, and water gurgling from a faucet.  As shown in Fig.~\ref{fig:emd-calculus}, the noise suppression algorithm improves the ROC-AUC of detecting calculus by 0.04, mainly because the direct path signal from the toothbrush is suppressed. For the detection of food impaction, the noise suppression module improves the performance more significantly by 0.09 ROC-AUC value, as it significantly suppresses both the direct path signal and other environmental noise. }

% As shown in Fig.~\ref{fig:emd-1time} and Fig.~\ref{fig:emd-5time}, our Noise Suppression module can significantly improve the detection accuracy. Further, \name\ improves the detection accuracy of calculus, in terms of AUC values, from 0.69 to 0.76, and from 0.75 to 0.84, for 1-time and 5-times accumulative measurements, respectively.

\begin{figure*}[t]
\centering
  % \hspace{0.4in}
  \begin{subfigure}[b]{0.19\linewidth}
  \centering
  % \hspace{1in}
    \includegraphics[width=0.85\textwidth]{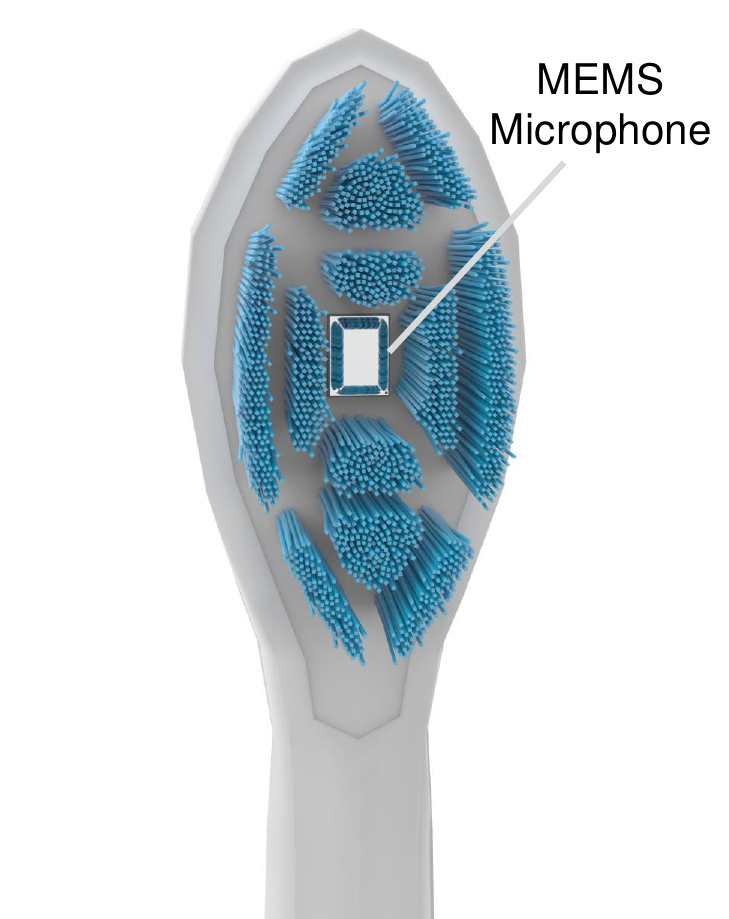}
    % \vspace{-0.2in}
    \caption{Prototype rendering}
    \label{fig:new-prototype}
  \end{subfigure}
  \hspace{0.05in}
  \begin{subfigure}[b]{0.25\linewidth}
    \includegraphics[width=\textwidth]{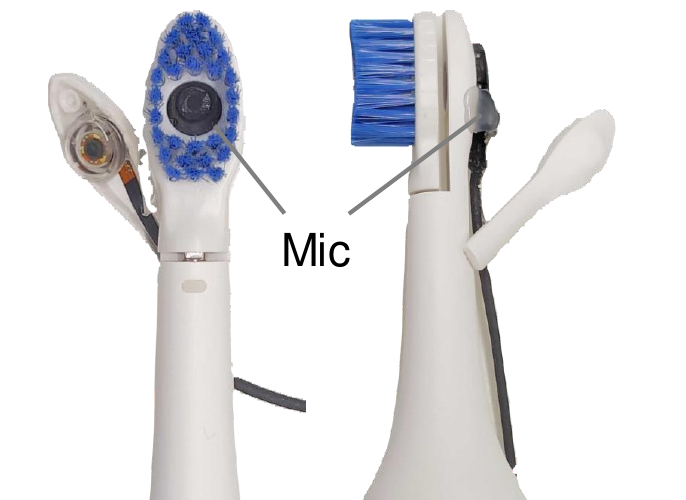}
    % \vspace{0in}
    \caption{Proof-of-concept Prototype}
    \label{fig:seebrush}
  \end{subfigure}
  \hspace{-0.05in}
  \begin{subfigure}[b]{0.28\linewidth}
    \includegraphics[width=\textwidth]{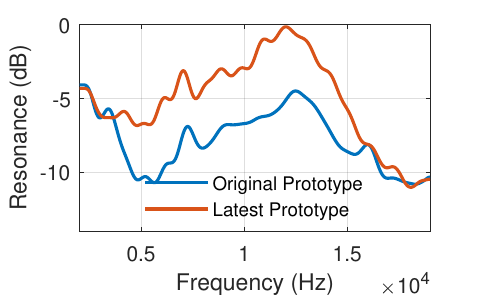}
    % \vspace{0in}
    \caption{Resonance before Damaged}
    \label{fig:next-gen-comparison}
  \end{subfigure}
  \hspace{-0.2in}
  \begin{subfigure}[b]{0.28\linewidth}
    \includegraphics[width=\textwidth]{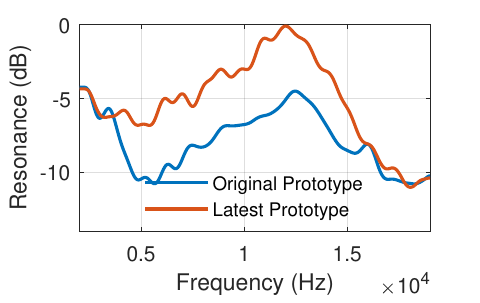}
    % \vspace{0in}
    \caption{Resonance after Damaged}
    \label{fig:next-gen-comparison-after}
  \end{subfigure}
  \vspace{-0.1in}
  \caption{\rr{A concept diagram of the next generation prototype of \name. (a) shows the rendering of the ideal prototype. (b) shows a proof-of-concept prototype by modifying SeeBrush with a microphone.} \feb{(c) and (d) show the resonance signal enhancement (harmonics removed by algorithms) by comparing with the original prototype shown in Fig.~\ref{fig:hardware}. The experiment is conducted on a tooth from a dental model before and after damage.}}
 % \vspace{-0.2in}
  \label{fig:next-gen}
\end{figure*}

\sssec{Sequence Alignment:} \kuang{ We conducted the user study on three participants along with the health detection experiments for interdental food impaction in a home setting. The user brushes the teeth one by one in a same quadrant, from the innermost molar to the outermost incisor. We collected measurements on the original teeth quadrant, the teeth quadrant with food impacted after eating, and the teeth quadrant after clearing out the food. We take 5 measurements in each stage (15 measurements in total from a teeth quadrant). We use the measurements from the original normal teeth as references and perform alignment on the others. A video camera is put in front of the user's mouth to capture the real-time ground truth locations of the toothbrush. A lamp with a narrow light beam to used to brighten up the video. Fig.~\ref{fig:data-labeling} shows an example of a frame in the recorded video. We use an online video labeling tool~\cite{Labelbox} to label the tooth number being brushed in each video frame.}

We evaluate the sequence alignment on 90 collected teeth scanning sequences from 6 teeth quadrants in total. The results are shown in Fig.~\ref{fig:location-accuracy} and Fig.~\ref{fig:location-error}. We compare the results with a baseline algorithm which assumes the teeth scanning is performed at a uniform speed, equally distributed time to each tooth. We present the alignment performance when the status of some of the teeth in the quadrant changes (i.e. food impaction in our experiments) and without any changes compared with the status of the references. When the tooth status remains the same as the references, \name\ achieves an alignment accuracy of 82\% and an average error of tooth number 0.18, while the baseline algorithm has 68\% and 0.33 respectively. When the status of some of the teeth changes, \name\ achieves a 78\% localization accuracy and 0.21 average error of tooth number. The baseline algorithm achieves 62\% and 0.40 respectively. These results show the DTW algorithm in \name\ outperforms the naive algorithm on sequence alignment of teeth scanning. Also, the alignment performance only degrades slightly when tooth status changes since the selected features for the sequence alignment algorithm are designed to be resilient to health condition changes.

Fig.~\ref{fig:loc-parts} further shows how the different types of teeth, like Incisors, Cuspids, Premolars, and Molars, affect the alignment error. The figure shows the mean error of detecting the tooth number that the user is currently brushing, and how it varies over different types of teeth. Since the premolars are generally smaller and have irregular non-flat shapes compared with other teeth, it introduces slightly higher alignment errors. 
% As the molars are hard to reach and brush properly, it is also challenging to align them. This also similar to the incisors and cuspids, as it's challenging to brush each of them separately given their size is usually smaller than the brush head itself.

\subsection{Expert Evalution}\label{sec:expert}
\kuang{
To evaluate \name\ from the expert's point of view, we interviewed 2 board-certified dentists (D1 \& D2). D1 has 30 years of practice with expertise in restorative dentistry, and D2 has 8 years of practice with expertise in oral radiology. We investigated the following questions:
\begin{enumerate}
    \item Is our current data collection procedure valid? What would be your recommendations to improve the data collection procedure for our system?
    \item What is your opinion of the current performance of our system?
    \item What are the potential benefits of integrating this technology into the current dental healthcare system? What are the main limitations and barriers?
\end{enumerate}
}
\sssec{Data collection procedure:} 
\kuang{Both dentists agreed that our data collection procedure is sufficient and valid. For the recommendations, D2 points out the importance of collecting data from caries evolving at different stages, which can be left as future work. D2 mentioned: "\textit{If they are really small caries, we don't care. If they're really big carries, we can actually see it and be fine. So this becomes very useful when we can detect those in-between carries. But that will be a much harder data collection.
}" Besides, D1 suggested exploring the prototype of toothbrush heads that are designed for actual detection rather than existing ones designed for hygiene.}

\sssec{System Performance:} \kuang{Both dentists stated that the overall detection accuracy of our system is promising, especially considering our current system is just using a regular toothbrush without specialized hardware. Meanwhile, both dentists agreed that, from a clinical point of view, it is necessary to further optimize the system and improve its accuracy. D2 mentioned: "\textit{I think the results are promising... When using cumulative 5 times measurements, the accuracy is consistently high... But we will have to tweak everything. The calculus one with one-time measurement seems to work at 0.72, which is not decent.}}"

\sssec{Benefits to current dental healthcare system and potential barriers:} \kuang{Both dentists considered that the potential of at-home early caries detection of our system is beneficial to the dental healthcare system, since it can catch caries earlier, prevent more severe diseases from happening and reduce the overall costs of the system. D1 also mentioned detecting unseen food impaction will have a direct impact on the patient's overall periodontal health. D2 emphasized the importance of at-home dental calculus detection: "\textit{We recommend the patient do come for a checkup every 6 months, but in practicality that is not possible for a lot of people. So having that calculus detection system will also again help the patient understand that there's some issue going on, and they need to go to the dentist and get it fixed before it becomes a much bigger issue, such as periodontal bone loss.}}"

\kuang{For the potential barriers, D2 is concerned about the accessibility of our system to low-income families. Families without access to regular dental visits may also lack powerful enough smartphones to receive and process the data. D1 is concerned about the interpretation of data by the patients and suggests the actual system can present the result by incorporating the dentist's interpretation.} 

\section{Discussion and Limitations}\label{sec:discussion}

%%%% tooth paste and liquide effect?
%%%% Users with already cavity can still use our system?
%%%% orientation change?

\sssec{Next generation prototype:} \rr{While our approaches and the proof-of-concept hardware setup (shown in Fig.~\ref{fig:hardware}) demonstrates promising results, future areas of exploration still remain. One may be concerned that our current setup integrates the microphone with a separate glass stick, which may induce inconvenience and instability for everyday use. In the future, we envision building a next-generation prototype of \name\ that integrates a tiny MEMS microphone into the center of the toothbrush head (as shown in the concept diagram in Fig.~\ref{fig:new-prototype}). Such design not only ensures usability, but also shortens the distance between the vibrating teeth and the microphone, which has the potential to improve the Signal-to-Noise Ratio~(SNR) and the overall sensing performance. }

\rr{To demonstrate the feasibility of the proposed prototype, we conduct a proof-of-concept experiment by modifying the SeeBrush~\cite{seebrush} (a camera-integrated toothbrush). As shown in Fig.~\ref{fig:seebrush}, we fixed the microphone to its original camera location on the toothbrush head. A preliminary experiment by brushing on a dental model shows promising results as in Fig.~\ref{fig:next-gen-comparison}. Comparing with our original setup (Fig.~\ref{fig:hardware}), this new prototype obtains very similar resonance behaviors when observing peaks and valley locations. Moreover, in the main resonance frequency range of the object (4-16~kHz), the new prototype achieves a mean value of 3.75~dB signal power enhancement as well as captures more detailed variations, mainly attributed to the shorter signal path from the tooth to the microphone.} \feb{We further damage the tooth using an electric drill and perform the measurement again. Similar changes in the resonance signal can be observed (e.g. the peak disappears at 7~kHz) for both the original prototype and the new prototype, which validates the potential of the new prototype for more convenient health sensing. For real deployment, we note that the resonance signatures obtained from different hardware are not comparable, i.e. the reference samples and testing samples should be obtained from a same hardware setup to ensure consistency.}

\sssec{Design choice of changes detection:} One may wonder why \name\ chooses to detect changes rather than detecting disease directly at once. We note that dental health sensing is a personalized task since the oral condition including the acoustic resonance of different people can vary widely. For example, one's healthy teeth may have a very similar resonance compared with another's teeth with caries. Building a normal discriminative model to judge the dental condition would require a large amount of labeled data and might not be accurate due to individual differences. Instead, \name\ chooses the approach that compares the new measurements with the previous references collected from the same person in a healthy state. This approach can be beneficial as it only focuses on detecting changes to achieve personalized sensing.

\sssec{Sensing during cleaning:} Though \name\ is based on off-the-shelf sonic toothbrushes, it requires an additional dedicated procedure to perform dental health sensing than regular oral hygiene and does not involve toothpaste. \kuang{The video-guided procedure instructs the user to brush their teeth one or two at a time, staying on the chewing surface of each for 3-4 seconds.}
% The designed teeth scanning procedure ensures that the teeth from the same quadrant are brushed in a fixed order, and the brushing procedure is performed from a relatively similar orientation. In this way, the sequence alignment algorithm can find the correct match for the signatures from the same tooth to enable further health detection. 
For future work, we aim to collect a larger dataset in the field and leverage data-driven approaches such as Connectionist Temporal Classification~\cite{ctc} based on deep learning, to support dental health sensing, despite an arbitrary brushing pattern.

\sssec{Long-term monitoring to detect different severity of conditions:} \kuang{The current user study of \name\ does not involve long-term monitoring to distinguish different severity of diseases (e.g. size of caries). However, the expert interviews pointed out the value of detecting caries at mid-stage. For future work, we aim to design and conduct long-term experiments incorporating dental clinics and patients, to validate and optimize the \name's detection performance for diseases at different stages.}

\sssec{Britle deterioration on toothbrush head:} \rr{One may be concerned that the captured tooth resonance signature would become inconsistent when the bristle on the toothbrush head undergoes wear and tear. To solve this problem, users can test the extracted signature of \name\ on a reference object with a wide resonance band (e.g. a model tooth) before performing a self-exam. If a clear difference can be observed in the signatures from the same reference object, it indicates that the toothbrush head needs to be replaced.}

\sssec{Multipath:} The audio signal generated by the tooth resonance propagates in the user's mouth taking both Line-of-Sight~(LOS) and non-Line-of-Sight~(nLOS) reflective paths. This causes potential interference to tooth resonance signatures. \name\ mitigates this problem by affixing the microphone as close as possible to the toothbrush (1.25~cm shown in Fig.~\ref{fig:hardware}), which makes the LOS path much more dominant than other nLOS paths. \rr{Our latest design shown in Fig.~\ref{fig:new-prototype} will further remedy this issue.} Besides, as \name\ only seeks to detect changes, we can make a reasonable assumption that the multipath changes inside one's mouth are relatively small even across days, which can be neglected. 
% For future work, we can integrate the microphone into the exact location of the brush head to further improve the signal quality.

% \sssec{Multipath:} The audio signal generated by the toothbrush propagates in the user's mouth taking different paths, which can affect the modeling of tooth resonance and diseases. This multipath effect may vary from a user to another, from environment to another, and more importantly from a tooth to another. \name\ attempts to minimize such issues by separately modeling each tooth, and detecting its health problems through differential measurements over time.

% \sssec{Teeth boundary:} Human teeth are usually well connected without in between spaces, specially for healthy teeth, which complicates alignment process and tooth localization. 

% \sssec{Smaller teeth:} The premolar and cuspid teeth of some users have a smaller size than the toothbrush head we chose. For these cases, we choose to model the signature derived from two or three teeth brushed together and perform health detection. We note that as a result, our system would provide information of health changes flagged across a small group of teeth rather than individual teeth, in these cases. 
\sssec{Other dental conditions:} We believe that \name\ can potentially be extended to detect dental conditions beyond those described in this paper. Technically, \name\ can sense any dental disease that causes changes to the teeth resonances across a wide frequency range. For example, teeth problems such as abscesses and gum diseases such as periodontitis can potentially be diagnosed by \name. We leave a more comprehensive evaluation of the potentials of sensing based on \name's platform for future work. 

\sssec{Braces, bridges and implants:} We note that \name\ was not comprehensively evaluated in the presence of braces, bridges, and implants. We acknowledge that these require more carefully targeted user studies that are challenging to mount for all types of dental attachments. In general, we believe that \name\ can detect changes in acoustic signatures of a wide range of surfaces that the brush comes in contact with. That said, braces may pose a unique challenge given that slight fluctuations in their position may also influence teeth signature. We leave a thorough investigation of this to future work. 
% \sssec{Sequence Alignment of teeth boundary:} 

\sssec{Incorporate other sensing modalities:} \kuang{As discussed in the related work section, multiple recent works propose to include sensing technologies into toothbrushing procedure, such as using IMU~\cite{mTeeth, mOral, luo2018brush, huang2016} or magnetic field~\cite{MET} for toothbrush tracking, as well as integrating blue-violet light camera on toothbrush head to monitor plaque removal~\cite{LumiO}. We believe the platform \name\ has great potential to incorporate these sensing modalities to support better toothbrush tracking as well as health sensing.}

% \sssec{IRB Ethics Statement: } We note that all our user studies were IRB-approved through a multi-university collaboration that involved a dental school and an engineering department. Clinical studies were performed in collaboration with dentists, in compliance with all safety and privacy standards. 

\section{Conclusion}
This paper presents \name, an in-house toothbrush-based solution for detecting common dental diseases such as caries, calculus and food impaction, and achieves an average detection ROC-AUC of 0.91, 0.83 and 0.88 respectively on a user study with 19 participants. \name\ achieves these objectives by disentangling the behavior of the teeth from the complex acoustic environment of the mouth and identifying important features that affect these conditions. There remain several open challenges for future researchers to build on \name\ to detect new diseases. We further believe future work can develop data-driven inference techniques to improve accuracy and new smart toothbrush designs that co-optimize sensing and cleaning.

\newpage
\bibliographystyle{ACM-Reference-Format}
\bibliography{reference}
\end{document}